\documentclass[twocolumn]{aastex631}

\usepackage{CJKutf8}

\newcommand{\dpn}[3]{$#1\,{\rm #2}_{#3}$}
\newcommand{\pf}[3]{$#1_{#3}^{\,#2}$}
\newcommand{\ppf}[3]{#1_{#3}^{\,#2}}

\newcommand{\appropto}{\mathrel{\vcenter{
  \offinterlineskip\halign{\hfil$##$\cr
    \propto\cr\noalign{\kern2pt}\sim\cr\noalign{\kern-2pt}}}}}
\defcitealias{Munoz21}{Paper I}
\shorttitle{Detailed Secular Evolution of Large TNOs}
\shortauthors{Mu\~noz-Guti\'errez et al.}
\begin{document}

%\title{Template \aastex Article with Examples: 
%v6.3.1\footnote{Released on March, 1st, 2021}}

\title{Long-term Dynamical Stability in the Outer Solar System. II. Detailed Secular Evolution of Four Large Regular and Resonant Trans-Neptunian Objects}

\correspondingauthor{Marco A. Mu\~noz-Guti\'errez}
\email{marco.munoz@uda.cl}

\author[0000-0002-0792-4332]{Marco A. Mu\~noz-Guti\'errez}
\affiliation{Instituto de Astronom\'ia y Ciencias Planetarias, Universidad de Atacama, Copayapu 485, Copiap\'o, Chile}

\author[0000-0001-7042-2207]{Antonio Peimbert}
\affiliation{Instituto de Astronom\'ia, Universidad Nacional Aut\'onoma de M\'exico, Apdo. postal 70-264, Ciudad Universitaria, M\'exico}

\author[0000-0002-5974-3998]{Angeles P\'erez-Villegas}
\affiliation{Instituto de Astronom\'ia, Universidad Nacional Aut\'onoma de M\'exico, A. P. 106, C.P. 22800, Ensenada, B.C., M\'exico}

%\author[0000-0003-4077-0985]{Matthew J. Lehner}
%\affiliation{Institute of Astronomy and Astrophysics, Academia Sinica, 11F of AS/NTU Astronomy-Mathematics Building, No.1, Sec. 4, Roosevelt Rd., Taipei 10617, Taiwan, R.O.C.}
%\affiliation{Department of Physics and Astronomy, University of Pennsylvania, 209 South 33rd Street, Philadelphia, PA 19125, USA}

%\author[0000-0001-6491-1901]{Shiang-Yu Wang (\begin{CJK*}{UTF8}{bkai}    王祥宇\end{CJK*})}
%\affiliation{Institute of Astronomy and Astrophysics, Academia Sinica, 11F of AS/NTU Astronomy-Mathematics Building, No.1, Sec. 4, Roosevelt Rd., Taipei 10617, Taiwan, R.O.C.}



\begin{abstract}

The long-term evolution of the outer Solar System is subject to the influence of the giant planets, however, perturbations from other massive bodies located in the region imprint secular signatures, that are discernible in long-term simulations. In this work, we performed an in-depth analysis of the evolution of massive objects Eris, \dpn{2015}{KH}{162}, Pluto, and \dpn{2010}{EK}{139} (a.k.a. Dziewanna), subject to perturbations from the giant planets and the 34 largest trans-Neptunian objects. We do this by analysing 200, 1 Gyr long simulations with identical initial conditions, but requiring the numerical integrator to take different time steps for each realization. Despite the integrator's robustness, each run's results are surprisingly different, showing the limitations of individual realizations when studying the trans-Neptunian region due to its intrinsic chaotic nature. For each object, we find orbital variables with well-defined oscillations and limits, and others with surprisingly large variances and seemingly erratic behaviors. We found that \dpn{2015}{KH}{162} is a non-resonant and very stable object that experiences only limited orbital excursions. Pluto is even more stable and we found a new underlying constraining mechanism for its orbit; \dpn{2010}{EK}{139} is not well trapped in the 7:2 mean motion resonance in the long-term and cannot be trapped simultaneously in von-Zeipel-Lidov-Kozai resonance; and finally, we found that at present Eris's longitude of perihelion is stationary, tightly librating around 190$^\circ$, but unexpectedly loses its confinement, drifting away after 150 Myr, suggesting a missing element in our model. 

\end{abstract}



\section{Introduction} 
\label{sec:intro}

The long-term dynamical evolution of the outer Solar System, in particular of the trans-Neptunian (TN) region, has been explored in varied ways, both numerically and analytically \citep[e.g.][]{Fernandez80,Duncan95,Malhotra00,Hahn03,Gallardo12,Saillenfest20}. The study of the TN region boasts importance since the orbital distribution of the small icy objects inhabiting this region carries the clues of a violent past through which the giant planets (GPs) went at the early stages of the formation of the Solar System \citep{Nesvorny18}.

Thanks mostly to these observed orbital properties of the trans-Neptunian objects (TNOs) a consistent picture has emerged, which describes how the four GPs, formed much closer to the Sun than we see them today, in a compact orbital configuration, migrated outwards due to their interaction with massive remnants of the planetesimal disk not accreted into planets \citep[e.g.][]{Fernandez84,Malhotra95,Ida00,Tsiganis05}. In this way, the orbits of Saturn, Uranus, and Neptune expanded their orbits to the current observed locations \citep[even possibly ejecting a 5th GP, an icy one similar to Uranus and Neptune, e.g.][]{Thommes99,Nesvorny12}. The scattered off planetesimals, on the other hand, are just a trace of the original planetesimal disk, the main bulk of which debris we observed today in the TN region\citep[e.g.][]{Morbidelli08,Nesvorny18,DeSousa20,Gladman21}. 

Though the characterization of the steady-state evolution of the outer Solar System (i.e. after the end of the migration phase of the GPs) has been explored in a systematically and statistically significant way \citep[see for example][]{Robutel01,Lykawka05,Munoz21,Forgacs23,Balaji23}, the particular evolution of individual objects has been mostly neglected, except for some objects of particular importance, such as the first-ever discovered TNOs (besides Pluto), Arrokoth, and of course Pluto itself \citep{Cohen65,Morbidelli95,Porter18,Malhotra22}.

The formation of new short-period comets shows that the TN region is not completely stable \citep[e.g.][]{Duncan88,Torbett89,Levison97,Emelyanenko04}. The creation of those low-inclination comets implies that over time many objects are continuously ejected from the TN region \citep[e.g.][]{Dones15,Nesvorny17}. Careful numerical studies show that for every new comet, there are a dozen or so TNOs ejected from the Solar System \citep[e.g.][]{Fernandez80,Munoz19}; this also shows that there is a very complex orbital evolution of those TNOs before being ejected from the Kuiper belt (KB). A similar complex evolution to that followed by small icy objects should be expected for more massive objects in the TN region such as dwarf planets (DPs); even a similar ejection rate (per object) cannot be discounted.

In the present work, we intend to contribute to increasing the inventory of well-characterized large TNOs, for which a long-term evolution of their orbits, as well as their statistical secular outcomes, is established by means of an ensemble of detailed numerical simulations of the outer solar system, which considers the perturbations form the GPs and the 34 most massive objects of the KB \citep{Munoz19,Munoz21}. Though the actual evolution of the ``real'' orbit is impossible to determine, from a large ensemble of long-term simulations, we can determine which orbital characteristics we can trust and which we cannot, since they depend strongly on the detailed interactions of each individual DP with the other massive objects of the KB.

This work is organized as follows. In Section \ref{sec:N-body} we present the description of the $N-$body simulations. Section \ref{sec:stable} shows the analysis of the stable objects Eris and \dpn{2015}{KH}{162}, while Section \ref{sec:resonant} includes the analysis of the resonant objects Pluto and \dpn{2010}{EK}{139}. Finally, in Section \ref{sec:conclusions}, the results are summarised and discussed.

\section{$N-$body Numerical Simulations
\label{sec:N-body}}

In previous works \citep{Munoz19,Munoz21}, we performed a total of 200 distinct simulations, in which the 34 largest TNOs in the Solar System were evolved for 1~Gyr, considering the gravitational influence of the Sun and the four GPs, as well as the mutual gravitational perturbations of the 34 TNOs themselves. In what follows, we will refer to all 34 objects as ``dwarf planets'' (DPs), even if the smallest ones might not be round (i.e. not true DPs). 

The simulations were started on the Julian Date 2458176.5 (February 27th, 2018), according to the data available at that time at the Minor Planet Center web page\footnote{\url{https://www.minorplanetcenter.net/iau/mpc.html}}. The initial conditions of our simulations in heliocentric Cartesian coordinates can be found as supplementary material. To integrate the orbits of each object, we used the hybrid Symplectic integrator from the {\sc MERCURY} package \citep{Chambers99}, with an accuracy parameter for the Bulirsch-Stoer integrator set to $10^{-10}$, and an ejection distance of 20,000 au. To explore the chaotic nature of the KB, we used these 200 integrations that originated from identical initial conditions but evolved with slightly different time-steps.

From our set of 200 total simulations, 177 simulations differ from each other by the number and distribution of the set of test particles included in them \citep[test particles were originally taken from the L7 model of the KB of][]{Petit11,Gladman12}, this set of 177 simulations was developed to study the evolution of cometary nuclei \citep{Munoz19}; the interaction with test particles results in random adjustments of the time-step size of the Bulirsch-Stoer integrator, which enters into play during close encounters, defined by a limit of 4 Hill radius around each GP and 5 around each DP. To complete our set, we performed 23 additional simulations where the initial time-steps were varied from 385 to 407 days. Though fewer perturbations are produced in the no-test particle simulations, we found no significant statistical differences among the subsets of simulations. Those subtle changes ensure the variability of the outcomes after 1 Gyr of integration of the different simulations despite the identical initial conditions of the massive objects. More details about the original simulations and our interpretation of the small random changes in time-step sizes as well as on the effects of the DP-DP close encounters can be found in \citet{Munoz19,Munoz21}.

\subsection{Long-term evolution of a sample of DPs}

In \citet[][hereafter \citetalias{Munoz21}]{Munoz21}, we classified our 34 DPs into three categories, according to the results of their 200 distinct realizations, namely stable, resonant, and unstable objects. In total, 17 objects were stable (or regular), 6 were resonant, located inside different mean motion resonances (MMRs) with Neptune, and 11 resulted in highly unstable or chaotic evolution. 

It must be noted that the regularity was only sought for the 3 main orbital parameters ($a$, $e$, and $i$). When looking into the characterizing angles ($\varpi$, $\Omega$, and $\lambda$), our simulations show that, within a few tens of millions of years, each realization will have circulated in a sufficiently different manner that all the information about the original phase will be lost. 

In this work, we show the detailed evolution of the selected members from the regular and resonant categories, some which we consider representative of the whole group. Besides their importance as referential objects in the KB (due naturally to their large masses), their characteristic evolution would serve as indicators that would permit the identification of other members into any of these two broad dynamical families; such new members would not necessarily be as massive and could be identified with a smaller ensemble of realizations (clones) by the similarity of their evolution on a long-term basis. This kind of qualitative classification is not uncommon, especially when detailed numerical simulations would be necessary for more formal classifications of recently discovered or otherwise interesting objects \citep[see for instance][]{Bannister17,Khain20}. However, the family of unstable objects (chaotic; see \citetalias{Munoz21}) shows a wide and complex variety of evolutionary tracks, it is not possible by their very nature to find repetitive or predictable behavior, thus we leave the detailed study of members of the unstable family for future work (Mu\~noz-Guti\'errez et al., in preparation). 

\section{Stable Objects
\label{sec:stable}}

The family of stable objects is the most populous in our sample, including 17 out of 34 objects (50\%). At the same time, this is the family among which is observed the smallest variation between individual orbits. This allows us to map out an allowed region of the orbital parameters and characterize their periodicity to some extent. Overall this makes the evolution of the objects of this family more predictable. 

The simulations ran in \citetalias{Munoz21} show that out of the 17 stable objects: 11 have all of their 200 orbits classified as regular (i.e. regular in all three orbital parameters), while the other 6 objects have each between 1 and 32 orbits classified as irregular. Overall there are a total of 59 orbits classified as irregular (out of 3400 in the entire stable family, i.e. 1.7\%). This irregularity is mainly due to changes in eccentricity (all the irregular orbits have changes in eccentricity larger than the empirical limit found in \citetalias{Munoz21}, $\left| \Delta e\right| > 0.0418$). Only a few orbits have important changes in the semi-major axis or inclination (3 of Eris's orbits have changes in $i$ and $a$, and only one of Makemake's orbits has changed in $a$).

\subsection{Eris}

%\begin{figure*}[ht!]
%\includegraphics[width=1.00\linewidth, height=9.0cm]{ERIS4panels.pdf}
%\plotone{ERIS4panels.pdf}
%\caption{Evolution of the orbital parameters of the 200 realizations of Eris in barycentric coordinates, see text for details. In the top three panels, we present the evolution of $a/a_0$, $e$, and $i$, respectively; in the fourth panel, we present $q$ in blue and $Q$ in red. The green dots highlight a single realization. \label{fig:Eris}}
%\end{figure*}

\begin{figure*}[ht!]
\includegraphics[width=\linewidth]{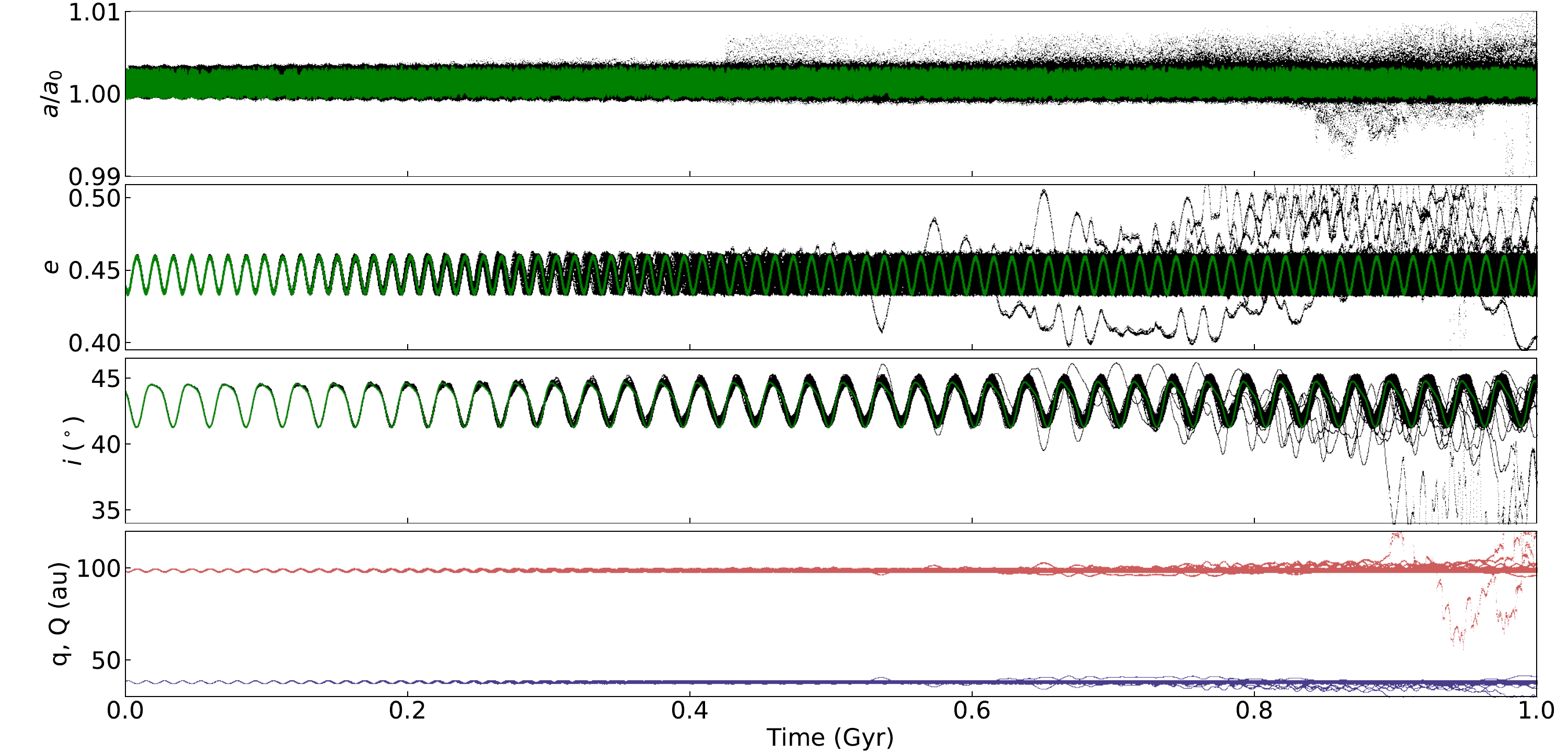}
\caption{Evolution of the main orbital parameters of the 200 realizations of Eris in barycentric coordinates, see text for details. In the top three panels, we present the evolution of $a/a_0$, $e$, and $i$, respectively; in the fourth panel, we present $q$ in blue and $Q$ in red. The green component highlights a single realization. \label{fig:Eris}}
\end{figure*}

%\begin{figure*}[ht!]
%\includegraphics[width=\linewidth]{ERIS_evol_zoom.pdf}
%\caption{Eris angles. \label{fig:Eris_zoom}}
%\end{figure*}

Being the most massive (as far as known) object in the TN region, Eris is widely considered stable. However, and despite being an object we classified as stable in \citetalias{Munoz21}, Eris has several close encounters with other DPs; due to these interactions, 9 of Eris's orbits end up leaving the phase-space region that we consider regular.

In Figure \ref{fig:Eris}, we present the 1 Gyr evolution of the orbital parameters of Eris: specifically, we show the evolution of the osculating orbital elements in barycentric coordinates, namely semimajor axis ($a/a_0$, where $a_0=67.87\,$au is the semimajor axis at the start of simulations), eccentricity ($e$), inclination ($i$), perihelion ($q$), and aphelion ($Q$); each point represents the values of the orbital parameters every 50,000 yr of integration. In the first 3 panels ($a/a_0$, $e$, and $i$) the points for all time steps of each of the 200 realizations of Eris's orbit are plotted as black dots; the area covered by those points defines a probabilistic region that the object could visit during its evolution; over these black points, we plot a single characteristic orbit in green to highlight the evolution of a typical case. In the fourth panel, we present $q$ as blue dots and $Q$ as red dots for all 200 realizations (without highlighting any individual orbit).

In the first panel of Fig. \ref{fig:Eris}, we can see a band that starts in the $0.9993<a/a_0<1.0033$ interval (i.e. $67.82 < a < 68.10\,$au) and becomes slightly wider towards the end of the integration, $0.9987<a/a_0<1.0040$ (i.e. $67.78 < a < 68.14\,$au). By analyzing Eris's evolution on any of its realizations, such as the green orbit, we found that there is a very fast oscillation (faster than 50,000 yr); to understand the origin of this oscillation we ran a single additional simulation for 1 Myr with much shorter time step (0.1 yr) and outputs (every 1,000 yr); we found that this fast oscillation is a complex response to the interaction of the 4 GPs, where the largest component of its power spectrum has a period of approximately 12,660 yr, which is impossible to analyze any further from our long-term simulations The effect of the fast oscillation is that in the top panel of Fig. \ref{fig:Eris}, each orbit appears like a small cloud with a $\sim 0.003 a/a_0$ width (i.e.$\sim 0.20\,$au), while the 200 realizations formed a dense band. An additional oscillation with a very small amplitude ($\sim \Delta a/a_0 < 0.0003$; i.e. $\sim \Delta a < 0.02 \, $au) and a period of about 13 Myr is also present (this oscillation can be observed more clearly in the second panel, corresponding to $e$). After the first 400 Myr, some orbits go out of the main band; by the end of the simulation, a total of 9 orbits have outed the main band: 6 of those orbits go out just slightly and form the small clouds above and below the main band; the remaining 3 orbits go out of the band by more than 3\%, however, they are not visible in this scale, which we decided to keep to have a better look at the 197 regular orbits in $a$.

In the second panel of Fig. \ref{fig:Eris} ($e$), we find that the average eccentricity or our 200 realizations is $e=0.4466$. The most outstanding feature in this panel corresponds to the $\sim$13 Myr oscillation, with an amplitude of 0.0122. This is the same period of oscillation that can be discerned, though only barely, in the top panel corresponding to $a/a_0$. We would expect also to have the fast variation identified in $a/a_0$; indeed, we can barely see a small thickening at the crests and valleys of the green orbit; overall the fast oscillations are about 9 times smaller than the 13 Myr oscillations with an amplitude of just 0.0014. During the first $\sim 250$ Myr all the 200 realizations oscillate more or less synchronized; with time, a noticeable fraction of the orbits shift the phase turning the black dots into a continuous ribbon, so by the 500 Myr mark, there is no longer a clear indication of a preferred phase. However, each of these orbits keeps a $\sim13$ Myr period, and overall the bulk of the orbits allow us somewhat to follow the associated oscillations on the top panel. As in the first panel, starting at $\sim 420$ Myr some of the realizations start to drift away from the main ribbon and, by the end of the simulations, 9 realizations will be irregular.

In the third panel of Fig. \ref{fig:Eris} ($i$) we show the behavior of Eris's inclination. By looking into the green curve, we see a main oscillation with a period of approximately 25.65 Myr and an amplitude of about $1.7^\circ$; we can also see that this oscillation is far from a sinusoidal. By looking more closely, we can see that the main oscillation is being perturbed by the 13 Myr oscillation that was seen in eccentricity; this second oscillation contributes with an amplitude of about $0.43^\circ$, but it is opposite to its value in eccentricity; since the period of the main oscillation is approximately double than that of the eccentricity oscillation, the main shape seems to continue for a while, but since the period ratio is not exactly an integer, there is a phase shift, and the shape of the sum changes to a more triangular shape by the end; this also changes the amplitude of the oscillations of the green line from $1.66^\circ$ at the beginning to $1.73^\circ$ near the end. If we subtract the 13 Myr component from the total oscillation, we find a well-defined sinusoidal with a period of 25.65 Myr and an amplitude of approximately $1.6^\circ$; this oscillation reflects a secular regression of Eris's line of nodes, more specifically a regression with respect to the Solar System's invariable plane. This $1.6^{\circ}$ angle is consistent with the $1.57^{\circ}$ inclination of the invariable plane with respect to the ecliptic \citep[see e.g.][]{Souami12}. 

It should be noted that though we use MERCURY's reference frame at the J2000 epoch, which results in a non-ideal frame of reference for the TN region \citep[see also][]{Huang22,Matheson23}, we decided to present our results with respect to the ecliptic to facilitate comparisons with other works. The invariability of both the frame of reference and the invariant plane translates into the precession of the orbital plane of Eris (and all the other objects studied in this paper) around the invariant plane resulting in a forced oscillation of $i$ that will have a fixed amplitude and remain present through the 1 Gyr integration. Since the invariant plane is determined by the total angular momentum, in principle our selection of DPs would technically impact its amplitude and orientation. However, it has been shown that there is a negligible contribution from dwarf planets (and all the other small body families) to the total angular momentum of the solar system of approximately 1'' in $i$ and 1' in $\Omega$ \citep[see e.g.][]{Li2019}.

The contribution of the fast oscillations found in $a$ and $e$ is negligible for $i$; also, the 25.65 Myr oscillation is not visible in either $a$ or $e$. Regarding the behavior of the few irregular orbits that stand out in the upper panels, we can see that the first disturbance appears at 540 Myr, and by 900 Myr all 9 irregular orbits are observable; six of them remain close to the main band, however, the other three go out of the band by reducing their inclination. 

Finally, in the fourth panel of Fig. \ref{fig:Eris}, we follow the perihelion ($q$, blue lines) and aphelion ($Q$, red lines) evolution of Eris; they can provide information about objects that get close to (or cross) Neptune's orbit. For both $q$ and $Q$, we see the oscillations related to the change in eccentricity, with the same 13 Myr period; the phase of these oscillations is completely lost around 400 Myr (at the same time as for $e$). We also observe the broadening of the bands after this time. Near the end of the integration, 3 of the irregular orbits get very close to Neptune, and one of them drastically reduces its aphelion from almost 100 au to around 60 au. Such drastic changes may result in a following chaotic trajectory or even an ejection, on longer timescales.

\begin{figure*}[ht!]
\includegraphics[width=\linewidth]{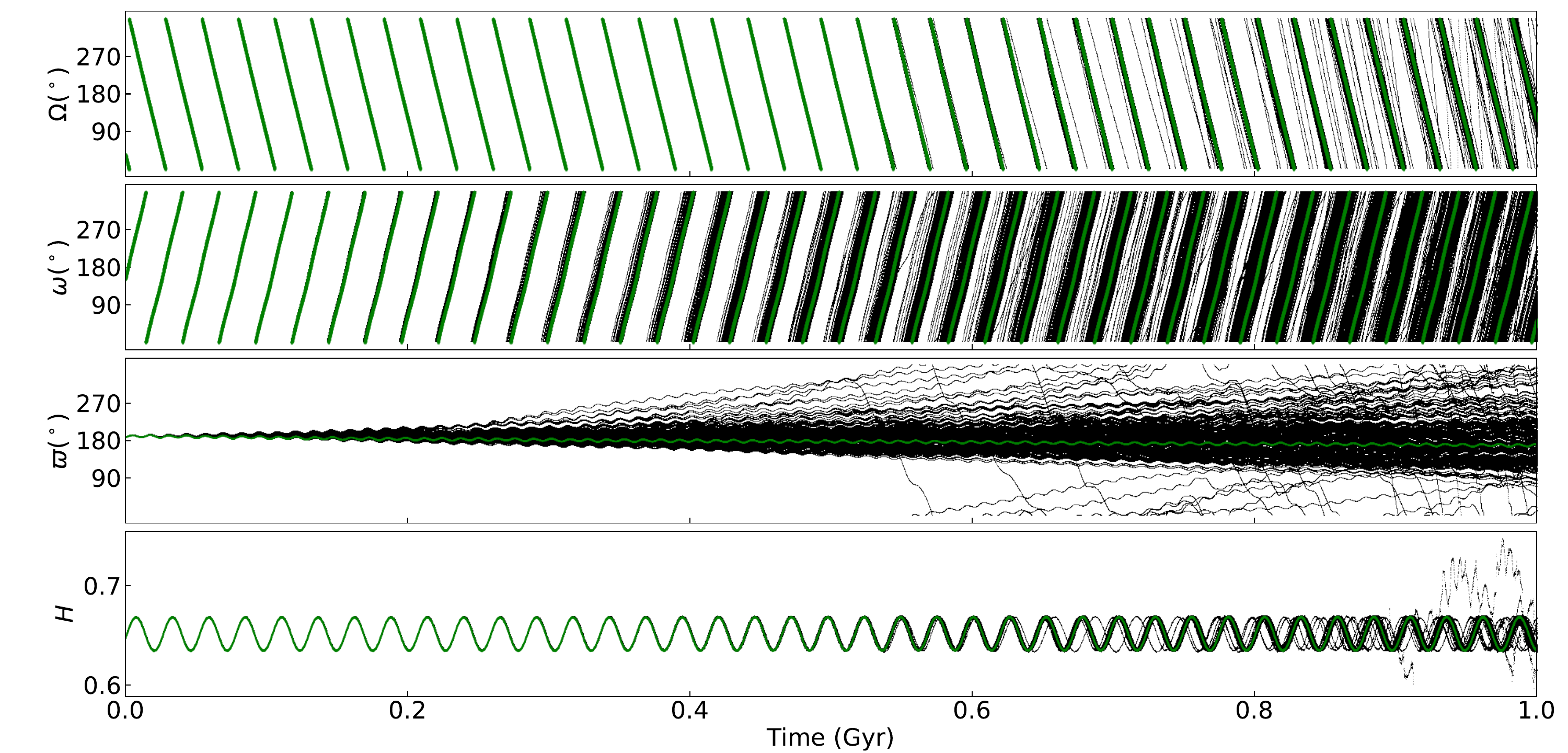}
\caption{Evolution of the angular elements of the 200 realizations of Eris in barycentric coordinates, see text for details. In the top three panels, we present the evolution of $\omega$, $\Omega$, and $\varpi$, respectively; in the fourth panel, we present the evolution of $H=\sqrt{1-e^2}\cos(i)$, Kozai's constant. As in Figure \ref{fig:Eris}, the green component highlights a single realization. \label{fig:Eris_angles}}
\end{figure*}

It is also possible to understand the long-term dynamics of Eris from an analytical point of view. Since Eris is located away from  Neptune's direct perturbations, i.e. with $q>38$ au, and outside any mean motion resonance (MMR; for Eris we found a diffusion time of approximately 65 Gyr, \citetalias{Munoz21}), its evolution will be only affected by secular perturbations, while due to its particular inclination it can be understood from the von-Zeipel-Lidov-Kozai dynamics \citep[vZLK, e.g.][and references therein]{Ito19}. According to the approximation developed by \citet{Gallardo12}, non-resonant secular evolution is governed by the following equations:
\begin{equation}
\frac{da}{dt}=0,
\label{eqn:a}
\end{equation}
\begin{equation}
\frac{de}{dt} = \frac{45ekE}{512a^{11/2}(1-e^2)^3} \left[5+7\cos(2i)\right]\sin^2(i)\sin(2\omega)+... \, ,
\label{eqn:e}
\end{equation}
\begin{equation}
\frac{di}{dt}=-\frac{45e^2kE}{1024a^{11/2}(1-e^2)^4}\left[5+7\cos(2i)\right]\sin(2i)\sin(2\omega)+... \, ,
\label{eqn:i}
\end{equation}
\begin{equation}
\frac{d\Omega}{dt}=-\frac{3Ck}{4a^{7/2}(1-e^2)^2}\cos(i)+... \, ,
\label{eqn:om}
\end{equation}
\begin{equation}
\frac{d\omega}{dt}=\frac{3Ck}{16a^{7/2}(1-e^2)^2}\left[3+5\cos(2i)\right]+... \, ,
\label{eqn:w}
\end{equation}
and
\begin{equation}
\frac{d\varpi}{dt}=\frac{3Ck}{16a^{7/2}(1-e^2)^2}\left[3-4\cos(i)+5\cos(2i)\right]+... \, ,
\label{eqn:varpi}
\end{equation}
where $k\approx0.017202~{\rm M}_\odot^{-1}~{\rm au}^{3/2}$ is the Gaussian gravitational constant, while $C$ and $E$ are the moments of inertia in the direction of $z$ for the secular model, which considers the planets as solid rings of mass $m_j$ with semimajor axes $a_j$. Considering the four GPs in circular orbits, those values are $C=0.1161$~M$_\odot$~au$^2$ and $E=52.0057$~M$_\odot$~au$^4$. 

From equation \ref{eqn:a}, $a$ is expected to be constant; in reality, the GPs are not rings and the true value of $a$ will shift slightly near a constant value (in our integrations, it remains within 0.27\% of a constant value: $a/a_0=1.0014\pm0.0027$). From equations \ref{eqn:e} and \ref{eqn:i}, it is possible to describe the interchange between $e$ and $i$ through Kozai's constant,
\begin{equation}
H=\sqrt{1-e^2}\cos(i);
\label{eq:H}
\end{equation}
again with fewer approximations or a more realistic scenario, $H$ is not expected to be exactly constant but to remain bound within a small interval. Finally, from equations \ref{eqn:om}, \ref{eqn:w}, and \ref{eqn:varpi} we can see, to first order, the sign of the evolution of $\Omega$, $\omega$, and $\varpi$: the longitude of the ascending node will constantly circulate backward for all prograde objects (regressing), the argument of perihelion will constantly circulate forward for all prograde objects (precessing), and the longitude of perihelion will depend only on $i$ (precessing when $i\lesssim46^\circ$ and regressing when $i\gtrsim46^\circ$). 

In Figure \ref{fig:Eris_angles}, we present the 1 Gyr evolution of these Eris angular elements in barycentric coordinates, namely longitude of ascending node, $\Omega$, argument of perihelion, $\omega$, longitude of perihelion, $\varpi$, and $H$, Kozai's constant; as in Figure \ref{fig:Eris} each point represents the values of the orbital parameters every 50,000 yr of integration for each of our 200 realizations; the area covered by those points defines a probabilistic region that the object could visit during its evolution; over the black points, we again plot a single characteristic orbit in green to highlight the evolution of a typical case.

In the top panel of Fig. \ref{fig:Eris_angles}, we can see that $\Omega$ is circulating backward constantly and more or less consistently across all realizations, as expected from equation \ref{eqn:om}; the period consistent with the 25.65 Myr found in $i$, and its phase consistent with maxima of $i$ at $\Omega=107.58^\circ$, i.e. is consistent with the longitude of ascending node of the invariable plane of the Solar System, found to be $107^{\circ}\, 34'\, 56''$ \citep{Souami12}. In the second panel, we see $\omega$ circulating forward consistently, as indicated by equation \ref{eqn:w}. For $\Omega$, the ensemble of realizations remains tightly bound for the entire integration, thus we can easily keep track of the number of circulations for all 191 regular realizations; while in $\omega$ the ensemble is not so tightly bound, after $\sim 500$ Myr a different number of circulations start to overlap thus by the end of the simulation, the coherence is lost even for the 191 regular orbits.

Although equation \ref{eqn:varpi} suggests that, to a first approximation, $d \varpi/ d t \approx 0$ when $i=46.378^\circ$; once we do the full integrations we find that, for Eris, the actual value for a non evolving $\varpi$ is closer to $i=43.2^\circ$, as can be seen in the third panel of Fig. \ref{fig:Eris_angles}; moreover, there is also a significant dependence on $e$, which produces a constant wave with an amplitude of approximately 3 degrees. Also of note is that the initial conditions for Eris place it in an orbit with very little evolution of $\varpi$; while this could be just a coincidence, this fine-tuning suggests its current values result from some resonance or stability process that guides it to these exact values of $a$, $e$, and $i$. Moreover, whatever the cause of this hypothetical process it is not part of our model, since within 150 Myr, most of the realizations have started to drift away from the equilibrium point.

Within the secular and non-resonant vZLK dynamics, $H$ is expected to be constant; in a full simulation we expect $H$ to be nearly constant --- at least when the orbits are in a stable regime. We can see in the bottom panel of Fig. \ref{fig:Eris_angles} that $H$ remains tightly bound within $0.651\pm0.017$; moreover, during the first 350 Myr all realizations remain tightly bound in a sinusoidal evolution with a period of 26.65 Myr, and without a hint of the 13 Myr oscillations observed in $e$ and $i$. For a very long time, $H$ remains oscillating with the same central value and nearly identical amplitudes; it lasts beyond 250 Myr when the ensemble of orbits starts losing cohesion in the other orbital and angular parameters, and even beyond 600 Myr when the cohesion of $H$ starts to unravel, it is only after 750 Myr that we find 3 of the orbits (those that are irregular in $i$) leaving the strip. 

The period of $H$ is the same as that of $i$, $\Omega$, and $\omega$. This is expected since, through equation \ref{eq:H}, H is linked to $i$, and thus its period is also identical to that of $\Omega$; and, in the beginning, it will also be identical to $\omega$, but they will never drift too much apart. Except for the three realizations that are irregular on $i$, for all the other realizations we can approximate $H$ by:
\begin{equation}
\label{eq:H_Eris}
H=0.017 \times \sin\left(\Omega - \Omega_{\it IPSS} \right)+0.651,
\end{equation}
where $\Omega_{\it SSIP}$ is the longitude of the ascending node of the invariable plane of the Solar System \citep[$\Omega_{\it SSIP}=107^{\circ}.58$;][]{Souami12}.

\begin{figure}[ht!]
\includegraphics[width=\linewidth]{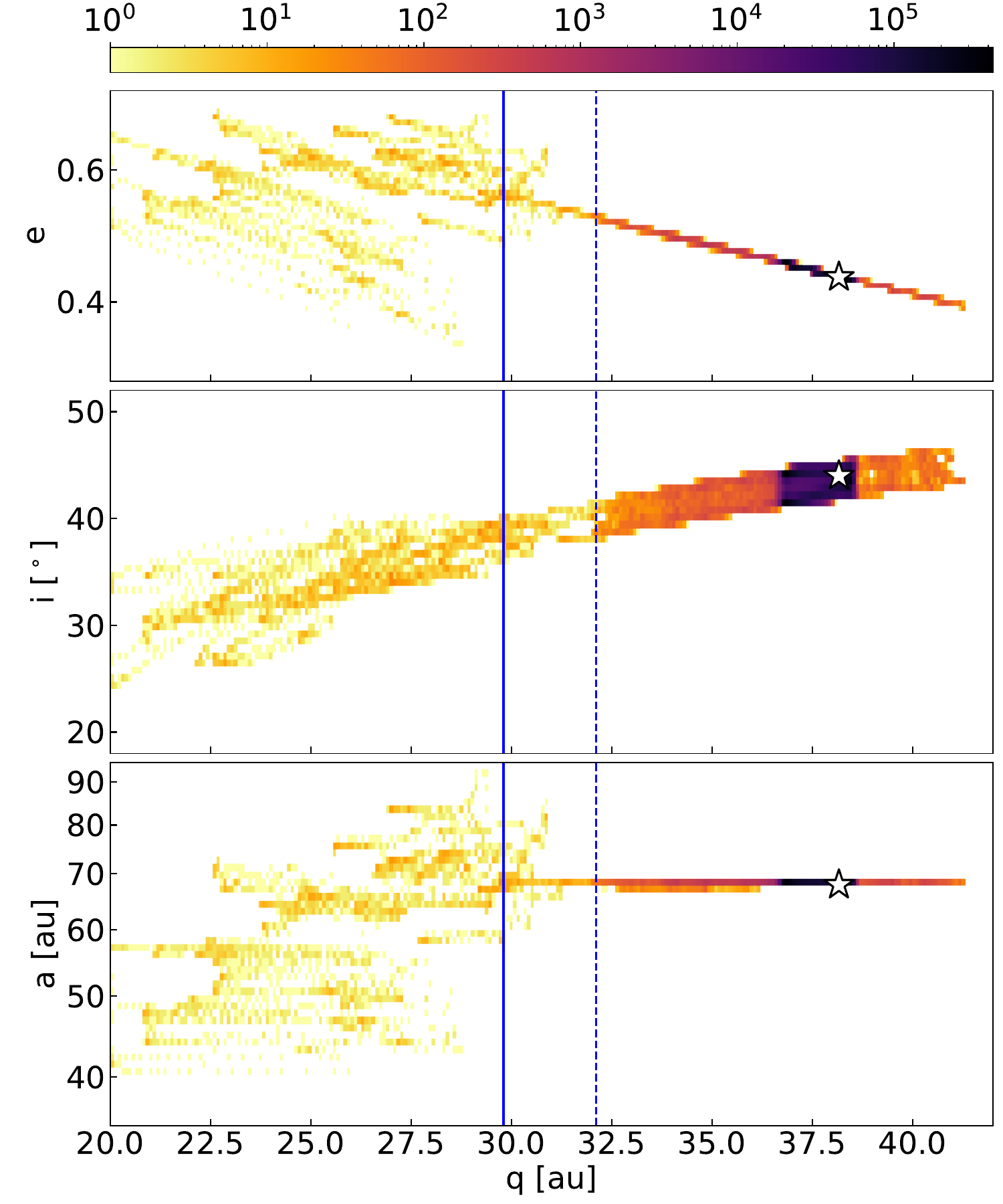}
\caption{Eris's orbital density, represented by a heat map based on the 200 realizations shown in Figure \ref{fig:Eris}. The white star indicates the initial condition. The color scale shows the number of points that fall into each bin across all simulations. All the regular orbits populated the dark area, the orange area with seven non-regular orbits, and two unstable realizations populated the yellow cloud at low $q$. The vertical blue lines are the location of Neptune's perihelion (solid) and the distance of 3 Hill radius away from Neptune's perihelion (dashed).\label{fig:Eris_density}}
\end{figure} 

To better visualize the global evolution of Eris, we finally construct density maps that measure the occupancy of the different Eris's realizations in different phase-space planes. These maps combine the data from all the simulations and permit us to understand Eris's stability as well as to locate the regions of phase space where it is likely to be found, on the one hand, but also Eris's possibility of becoming unstable (in limited cases), on the other.

In Fig.~\ref{fig:Eris_density}, we divided the space covered by the plots with a homogeneous grid in the linear space for the planes of perihelion vs.~eccentricity (top panel) and perihelion vs.~inclination (middle panel), and with a log-linear grid for the plane of perihelion vs. semimajor axis (bottom panel), counting the number of apparitions of an Eris data point along the 1 Gyr simulations, joining the data from the 200 different realizations. In this sense, the density map shows the statistical probability of Eris spending a determined amount of time in the respective area of the phase space, given by the density of points which are translated to a logarithmic color scale. The white star in all the panels represents the initial conditions of the simulations.

Looking at Fig.~\ref{fig:Eris_density}, one can immediately see the interchange between eccentricity and inclination driven by the vZLK mechanism. The same eccentricity variations, while keeping an almost constant semimajor axis, result in limited excursions of Eris's perihelion, i.e. the purple/black area ($36.7<q<38.7$~au) represents all 191 regular orbits as well as a significant fraction of each of the 9 irregular orbits, the orange and yellow colors represent the irregular part of the 9 irregular orbits that remain almost constrained (orange; $32.1<q<41.3$~au) until 2 realizations reach within 3 Hill radii of Neptune where they are no longer constrained (yellow). When the orbit of Eris approaches that of Neptune at least to 3 Hill radius of the giant (identified by the dashed blue line), the strong gravitational perturbations of Neptune result in the instabilities observed in the above Figures and the large excursions present in this plot. 

\subsection{\dpn{2015}{KH}{162}}

\begin{figure*}[ht!]
\includegraphics[width=\linewidth]{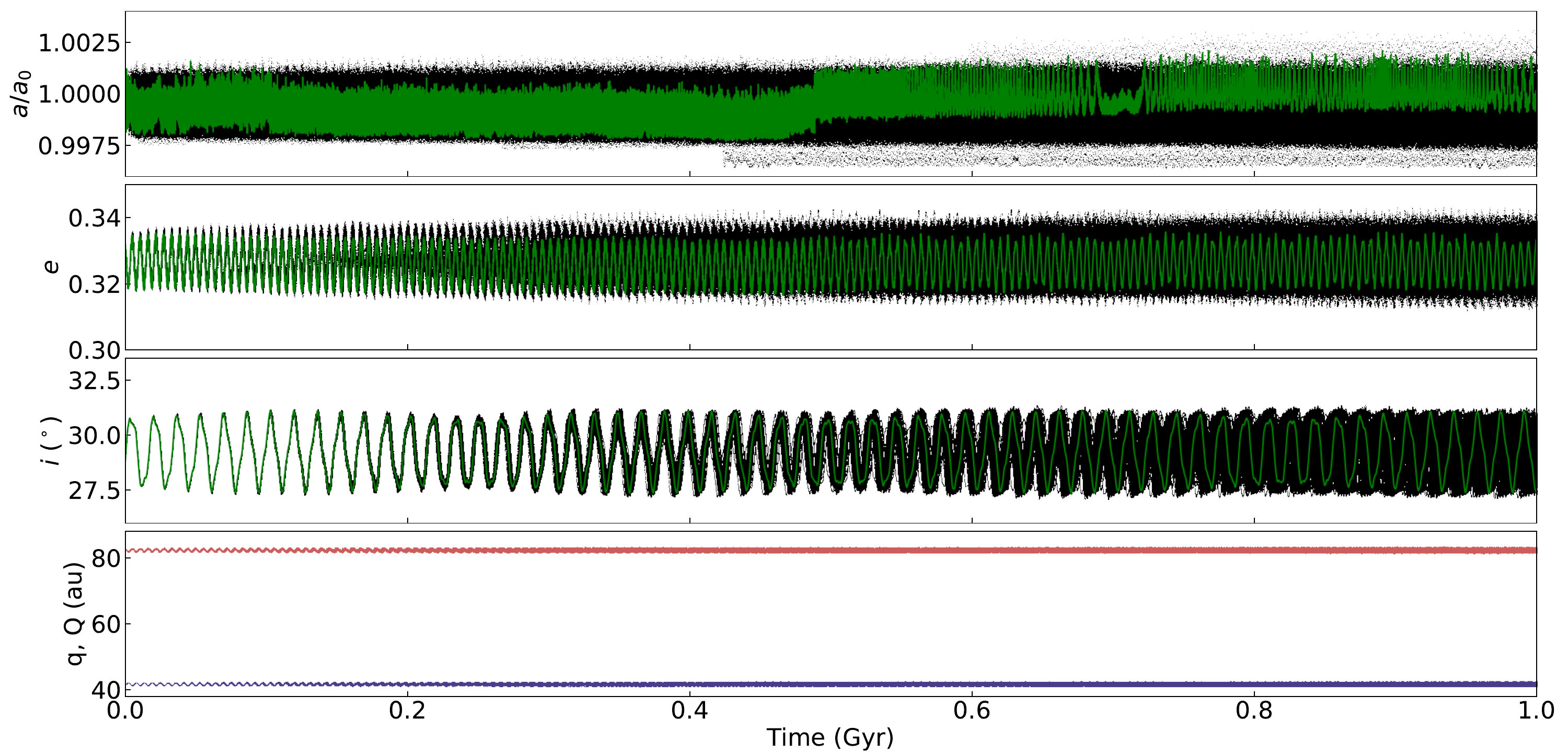}
\caption{Same as Fig. \ref{fig:Eris}, but for the evolution of the orbital parameters of \dpn{2015}{KH}{162}. \label{fig:2015KH16}}
\end{figure*}

\begin{figure}[ht!]
\includegraphics[width=\linewidth]{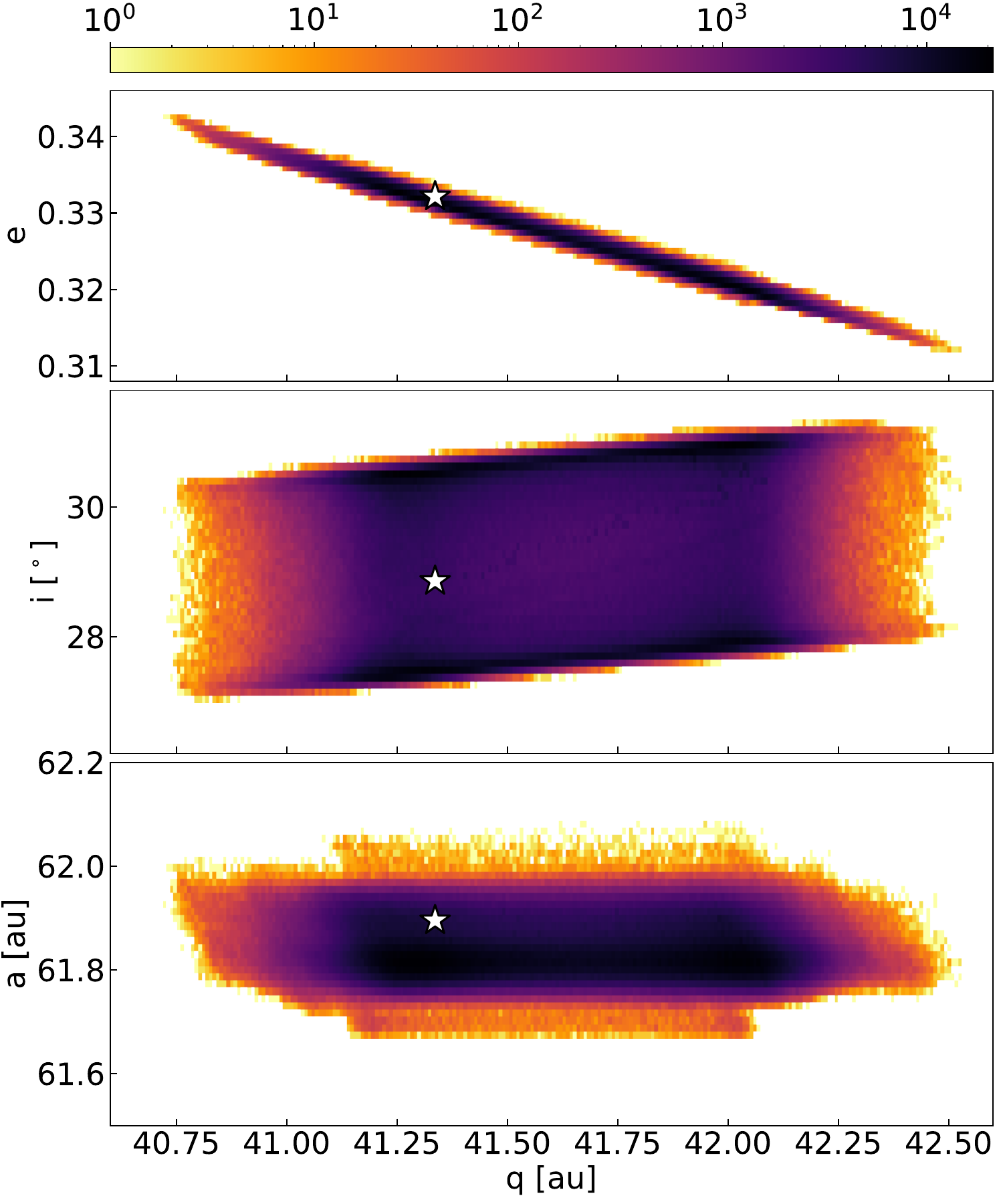}
\caption{Same as Fig. \ref{fig:Eris_density} but for the orbital density of \dpn{2015}{KH}{162}. \label{fig:2015KH162_density}}
\end{figure}

\dpn{2015}{KH}{162} was originally discovered by \citet{Sheppard16} and initially classified as a resonant object. Its most notable feature is its proximity to Neptune's 3:1 MMR and how such proximity impacts (or has impacted) its evolution. 

In the diffusion maps of \citetalias{Munoz21}, we found that \dpn{2015}{KH}{162} is indeed apparently located in the middle of the 3:1 resonance with Neptune; this happens if we consider barycentric coordinates. However, we were unable to find a single resonant argument of the 3:1 resonance that showed librating behavior, either for $ee'$ or for $II'$ type resonances. Also, all of our 200 realizations were found to be regular, making \dpn{2015}{KH}{162} an extremely stable object. Indeed, \cite{Kaib16} consider \dpn{2015}{KH}{162} as an example of a family of objects decoupled from Neptune's MMRs after planetary migration, when the vZLK mechanism raised the perihelion, decoupling it from Neptune and throwing it out of resonance. Also \citet{Saillenfest17} fail to find a librating resonant argument for \dpn{2015}{KH}{162}, thus we can conclude this object lies close to the resonance without actually being resonant. More recently, \citet{Alves23} studied the stability of the phase-space both inside and near the 3:1 MMR using frequency analysis, finding that the region surrounding the 3:1 MMR librating islands are almost as stable as the resonant regions, at least for eccentricities below $\sim 0.5$ in the presence of the four GPs.

Figure \ref{fig:2015KH16} is analogous to Fig.~\ref{fig:Eris}; here we present the 1 Gyr evolution of the barycentric orbital parameters of \dpn{2015}{KH}{162}. The top panel shows the evolution of $a/a_0$, where for this object $a_0=61.88\,$au. With a similar semimajor axis but smaller eccentricity than Eris, \dpn{2015}{KH}{162}'s perihelion never gets close enough to Neptune to have a close encounter with the GP, resulting in no extreme perturbations for any of the realizations of \dpn{2015}{KH}{162}. Overall, within the black band, we can see at least 2 encounters with other DPs, which do not result in perturbations large enough to affect the regular classification of any of the 200 orbits into the irregular regime; this kind of weak perturbations are easier to see in individual orbits, such as the two weak encounters within the green orbit. As with Eris, we find three oscillation patterns within the orbits, the fast one showing up clearly, with an amplitude of $\sim 0.0008$, in $a/a_0$ (an amplitude of $\sim 0.05$au). Note the apparent slowing down of the oscillations at around 700 Myr is an artifact of our sampling since the value of $a$ oscillates hundreds of times between each of our output points; it is not feasible to plot enough data points to try sampling this oscillation.

In the second panel of Figure \ref{fig:2015KH16}, we present the evolution of $e$. It is clear that the oscillations in eccentricity have a smaller amplitude but occur about two and a half times faster than in Eris; with an amplitude of $\sim 0.007$ in $e$ and a period of $\sim 5.6$~Myr. On top of this main $e$ oscillation, the same kind of rapid oscillation seen on Eris's eccentricity, with a period of thousands of years, is also present; zooming into the figure, the amplitude of this oscillation is $\sim 0.0015$. While all the realizations behave similarly, the coherence of the individual realizations starts to fade after $\sim250$~Myr.

The inclination evolution of \dpn{2015}{KH}{162} is shown in the third panel of Figure \ref{fig:2015KH16}. The main $i$ oscillation period is $\sim 16.4$~Myr, nearly two times faster than for Eris, and the amplitude is approximately $1.55^{\circ}$ again consistent with the $1.57^{\circ}$ present-day inclination of the Solar System's invariant plane with respect to the ecliptic. A second oscillation, with the same period as the main $e$ oscillation, is also found superimposed over the main oscillation of $i$; this oscillation is nearly three times faster than the main oscillation, producing some ``triangular'' shapes seen clearly at around 80~Myr, but its actual period is closer to 44/15 times faster, thus also creating the ``square'' shapes near 250~Myr. Coherence of the individual realizations does not last the entire integration and starts to fade after $\sim700$~Myr.

In the final panel of Figure \ref{fig:2015KH16}, we present the evolution of $q$ and $Q$ for \dpn{2015}{KH}{162}. The only feature of note, the small $\sim5.6$~Myr oscillation present in both curves, comes directly from the evolution of $e$ since $a$ remains almost unchanged. These oscillations in $q$ and $Q$ show the same period as well as the coherence of the ensemble of realizations as observed in $e$.

We present the orbital density map for \dpn{2015}{KH}{162} in Figure \ref{fig:2015KH162_density}. Since none of the 200 realizations is unstable, here we only see the equivalent of the black/purple core observed for Eris in Fig. \ref{fig:Eris_density}. However, we choose to zoom in as much as possible to make optimum use of the space for the figure. From the three panels in Fig. \ref{fig:2015KH162_density} we can easily determine the maximum excursions of \dpn{2015}{KH}{162} in all its orbital parameters, finding $\Delta_e\approx0.03$, $\Delta_i\approx4^\circ$, $\Delta_a\approx0.4$~au, and $\Delta_q\approx1.75$~au. These small scatters reflect the overall stability of \dpn{2015}{KH}{162}.

\section{Resonant Objects
\label{sec:resonant}}

The resonant family includes only 6 out of the 34 objects studied in \citetalias{Munoz21}, being the less numerous family in our sample. Those objects are located in the 3:2 (Pluto, Ixion, \dpn{2003}{AZ}{84}), 5:2 (\dpn{2002}{TC}{302}), 7:2 (\dpn{2010}{EK}{139}), and 21:5 (\dpn{2010}{JO}{179}) MMRs with Neptune and are all long-term stable. Yet, resonant perturbations result, in general, in a more interesting behavior than those of the stable family. 

Overall, out of 1200 total realizations (corresponding to 200 orbits for each one of the 6 resonant objects), 622 (52\%) are unstable in $e$, $i$, or both; however, only 17 (1.4\%) are unstable in $a$, i.e. 98.6\% of the orbits remain well within their resonances for the entire Gyr of integration.

Resonant objects can show large variations of the orbital elements without becoming unstable, unlike objects of the unstable family (see, for example, Fig. 7 of \citetalias{Munoz21}). From the set of 6 resonant objects identified in \citetalias{Munoz21}, we chose Pluto and \dpn{2010}{EK}{139} as the two representative resonant objects to analyze here in detail. 

The choice of Pluto seems natural given that it is the largest object in our resonant sample. Also, Pluto has received considerable attention for a long time making it the best-characterized object in resonance \citep[see e.g.][]{Cohen65,Williams71,Nacozy78,Milani89,Kinoshita96,Wan01,Malhotra22}. It must be noted that although Pluto is also chaotic \citep[having a positive Lyapunov exponent; see e.g.][]{Sussman88}, it is not unstable, thus we consider it a member of the resonant family and will discuss its weak chaotic nature further in section \ref{sec:pluto}. 

On the other hand, \dpn{2010}{EK}{139} shows a broad and interesting evolution of its orbital parameters, with only 2 realizations remaining stable in $e$ and $i$, while 198 realizations remain within its resonance. Furthermore, \dpn{2010}{EK}{139} has not yet been identified as resonant. We will describe its orbital behavior in section \ref{sec:EK139}.

\subsection{Pluto}
\label{sec:pluto}

\begin{figure*}[ht!]
\includegraphics[width=\linewidth]{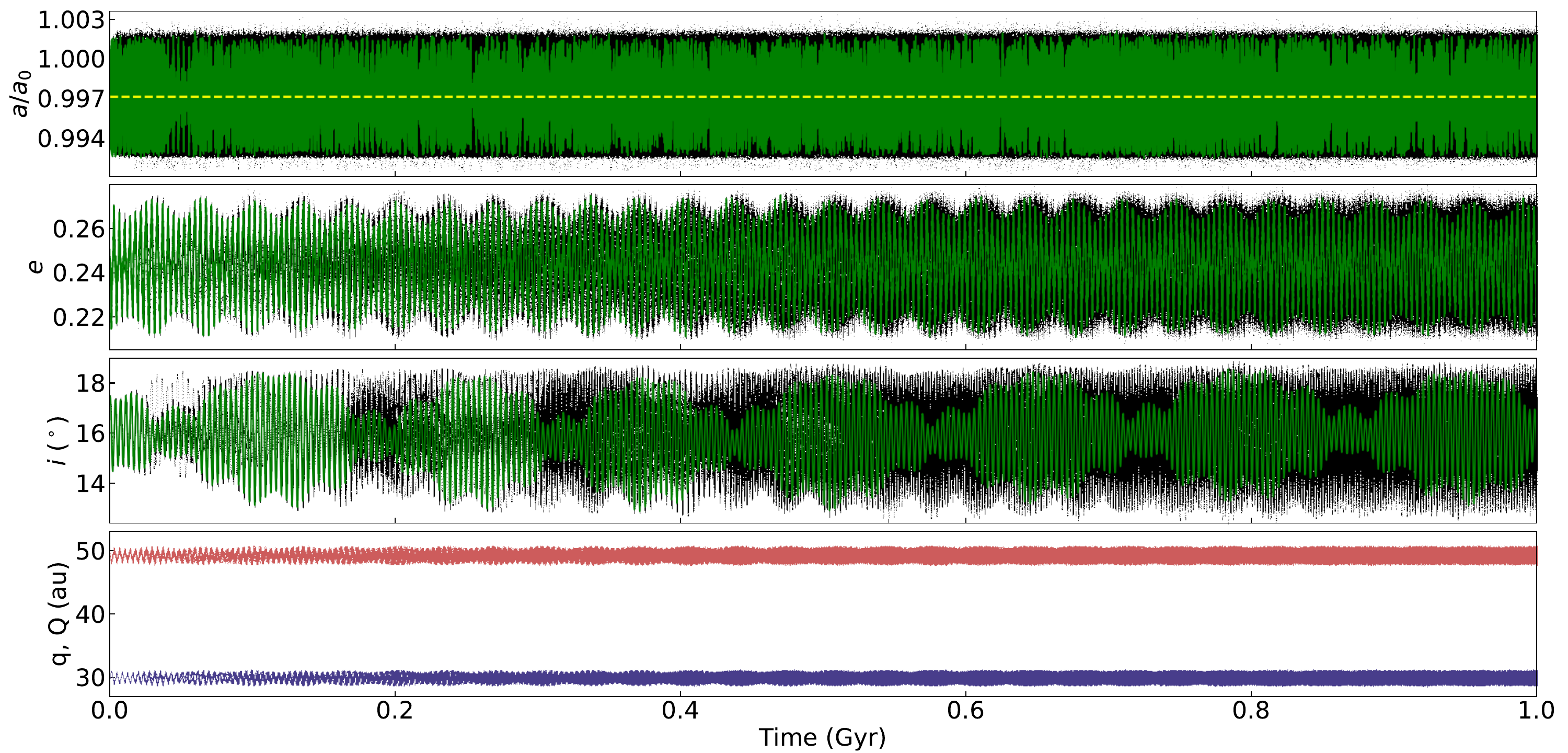}
\caption{Same as Fig. \ref{fig:Eris}, but for the evolution of the orbital parameters of Pluto. The dashed yellow line in the top panel represents the nominal location of the 3:2 MMR. \label{fig:Pluto}}
\end{figure*}

As mentioned before, Pluto has been detected as chaotic by determining a positive Lyapunov exponent in numerical simulations \citep[e.g.][]{Applegate86,Sussman88,Kinoshita96}, besides, there is a noticeable chaotic region in the $a$ vs $e$ plane just outside the 3:2 MMR with Neptune (see e.g. Fig. 3 in \citetalias{Munoz21}). However, Pluto is deeply locked within the 3:2 MMR and the expected chaos is not detectable in its main orbital parameters ($a$, $e$, and $i$). Indeed, this chaos does not lead to significant variations even on Gyr time scales, equivalent to many of Pluto's Lyapunov times \citep[where the Lyapunov time is $\sim$20~Myr, e.g.][]{Ito02}. 

%\marco{Este texto podria modificarse ligeramente en vista de papers mas antiguos que parecen mencionar la misma idea. Pero tengo que verlo con mas atencion.}
It has been recently suggested that Pluto is, in reality, only at the edge of the chaotic zone \citep{Malhotra22}, which could explain the apparent confinement of the chaotic effects expected from a positive Lyapunov exponent, as has been observed for other objects in the Solar System\citep[e.g.][]{Milani92}.

From the previous considerations and the results from \citetalias{Munoz21}, we conclude that Pluto either lies very close to a chaotic region or within a thin chaotic region with confined effects over the main orbital parameters. Either way, we can consider Pluto as a resonant dwarf planet, and we will study it accordingly.

A single realization of Pluto's orbit will show its quasi-periodic oscillations. However, the small numerical variations, introduced by the different step sizes of our integrations, can exponentially grow, due to the chaotic nature of Pluto's orbit; such growth allows them to fill a large part of the phase space determined by the Sun and the 4 GPs for Pluto to roam and an ensemble of initially identical orbits allows us to explore it. Beyond this chaos, the relevance of the available phase space is boosted by the presence of additional massive bodies able to perturb Pluto away from a theoretical “ideal orbit”.

\begin{figure*}[ht!]
\includegraphics[width=\linewidth]{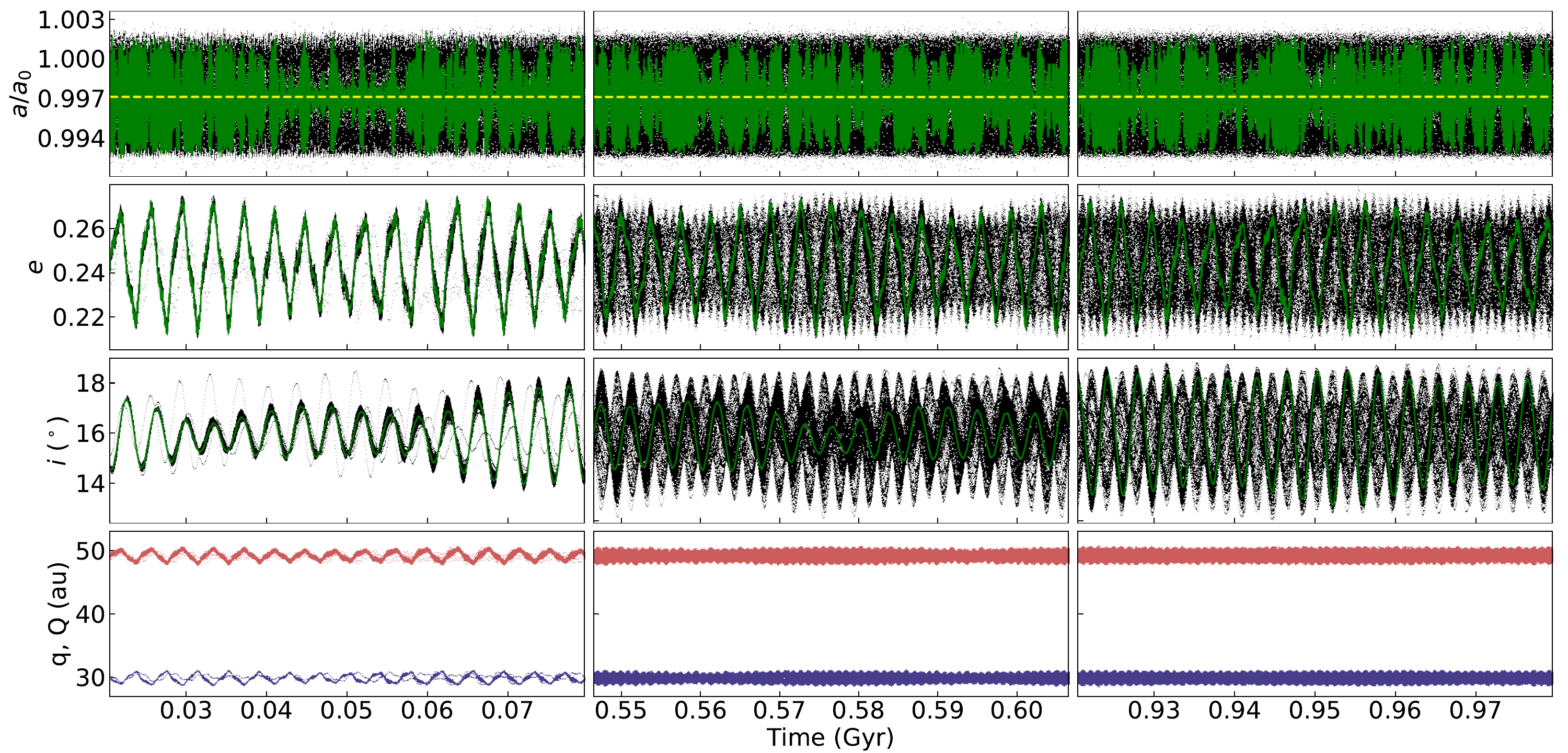}
\caption{Due to the rapid oscillations of $e$ and $i$, we zoom in to three selected shorter time intervals within Fig. \ref{fig:Pluto}; see text for details. \label{fig:Pluto_evol}}
\end{figure*}

In figure \ref{fig:Pluto}, we present the evolution of Pluto's $a$, $e$, $i$, $q$, and $Q$, similarly as we did for Eris and \dpn{2015}{KH}{162} in Figs. \ref{fig:Eris} and \ref{fig:2015KH16}, respectively. However, since Pluto is closer to the Sun, the oscillations of the orbital parameters evolve faster. Thus, in figure \ref{fig:Pluto_evol} we zoom in on the time axis and present only 180 Myr, divided into three parts, to see both: details of individual realizations and the ensemble behavior at different stages along the integration. The three columns of Fig. \ref{fig:Pluto_evol} represent stages near the beginning (left column, $20-80$ Myr), middle (center column, $546.5-606.5$ Myr), and end (right column, $920-980$ Myr) of our integrations. We selected these specific periods to highlight some features within the evolution of the oscillations in $e$ and $i$. 

In the upper row of panels of Figs. \ref{fig:Pluto} and \ref{fig:Pluto_evol}, we see the evolution of $a/a_0$, where $a_0=39.56$ au. Again, the evolution of the semimajor axis is much faster than our output cadence, so the ``speech-like'' pattern is only an artifact of the outputs, as each individual orbit fills the entire space between 0.993 and 1.001. It should also be noted that at 20 Myr the ensemble has already lost coherence, it is lost somewhere between 5 and 12 Myr. In contrast to both Eris and \dpn{2015}{KH}{162}, the evolution of each orbit is symmetrical around $a/a_0\approx0.9970$; which slightly deviates from the nominal location of the 3:2 MMR, $a_{3:2}/a_0=0.99715$, where $a_{3:2}=\left<{a_{Nep}}\right>(3/2)^{2/3}$ and $\left<{a_{Nep}}\right>=30.10411$ au is the average of all Neptune's semimajor axis values considering its 200 realizations; the yellow dashed line shows this resonance center; note that $\left<a_{Nep}\right>$ is about 0.016\% smaller than the current $a_{Nep}$. Also, contributing to the symmetric appearance of the green shadow is that the frequency of this oscillation is $\approx (2n+1)/100,000\;$yr$^{-1}$, with an unknown number of cycles (2n+1), producing a phase shift of $\approx 180^\circ$ between our 50,000 yr output steps. Note that with our output cadence, we cannot resolve the well-known oscillation period of the resonant argument of $\sim19,800$ yr \citep[e.g.][]{Cohen65,Williams71}, which would produce $2n+1\approx5$ oscillations for every 2 of our time steps. Finally, in contrast with the regular objects, for the entire 1~Gyr integration there is no clear change in the behavior of either the highlighted orbit or the ensemble: there is no evidence of a widening of the black ribbon across the integration or small shifts on $a/a_0$ due to close encounters with other DPs; this is due to the restoring effect provided by the resonance.

The second row of panels of Figs. \ref{fig:Pluto} and \ref{fig:Pluto_evol} presents the evolution of $e$. There we can see a complex pattern of oscillations that show at least 3 different periods: a) The main oscillation period in $e$ (\pf{P}{e}{o}$=3.795 \pm 0.002$Myr) that can be seen all across the green curve in Figs. \ref{fig:Pluto} and \ref{fig:Pluto_evol}. b) Considering the full ensemble there is an additional frequency exactly 4 times faster than the one shown by the green curve, the pattern period (\pf{P}{e}{p}$=0.9489 \pm 0.0005$Myr); in the 2nd and 3rd columns of $e$ in Fig. \ref{fig:Pluto_evol} we can see 3 black maxima perfectly spaced between each pair of green maxima (or minima), with a fourth black maximum behind each green one with \pf{P}{e}{o}$=$\pf{P}{e}{p}. c) While the third period corresponds to the modulation of the amplitude (\pf{P}{e}{m}$=34.4 \pm 0.1$ Myr), more clearly seen on the second panel of Fig. \ref{fig:Pluto}. 

While \pf{P}{e}{o} and \pf{P}{e}{m} are well known from the literature \citep[e.g.][]{Milani89,Kinoshita96,Malhotra97}, the existence of \pf{P}{e}{p} was previously unknown and quite unexpected. In the presence of small perturbations from other DPs, we would expect a random change in the amplitude and/or the phase of the oscillations, slowly driving the ensemble to lose coherence; instead, after each perturbation, we find two peculiar behaviors: a) the amplitude of the oscillations reverts to the original values in less than 10~Myr and b) if there is a phase shift, it will quickly settle down and be forcefully reorganized into the phases allowed by \pf{P}{e}{p}. While the first of these effects can easily be attributed to the strong perturbations of Neptune's 3:2 MMR; the second cannot be explained so easily. 
%We will return to this frequency after discussing $i$ and $\omega$.
We surmise this effect could be directly related to the orbital and physical configuration of the four GPs of the Solar System, which results in the libration of $\omega$ (reminding that without the giant planets, Pluto's argument of perihelion will not librate), therefore on the oscillation of both $e$ and $i$ with almost the same period, as we will discuss below. Note that \pf{P}{e}{p} is exactly a fourth of the known frequency \pf{P}{e}{o}, thus the underlying mechanism of both frequencies must be the same. However, the full explanation of the newly observed quantized perturbations would require a much more in-depth study, which is beyond the scope of this paper. 

%Note that \pf{P}{e}{p} is exactly a fourth of the known frequency \pf{P}{e}{o}, thus the underlying mechanism of this new frequency must have the same origin, indicating that the origin of the frequency \pf{P}{e}{o} is not well understood as was previously thought; besides aaaaa and bbbb, there is probably an additional dependence on the configuration of the 4 GPs that forces Pluto into having the \pf{P}{e}{o}) period.

On the other hand, \pf{P}{e}{m} is not bound by the previous effect and, by the end of the integrations, the modulation of the amplitudes presents a small phase drift across the ensemble.

The third row of panels of Figs. \ref{fig:Pluto} and \ref{fig:Pluto_evol} present the evolution of $i$. Here the oscillations have a more complex pattern than in $e$. We can detect the presence of two frequencies observed in $e$ (\pf{P}{e}{m}, \pf{P}{e}{o}) plus three new frequencies. Of the $e$ frequencies, the most obvious one is the pattern created by the modulation of the amplitude of $i$, as \pf{P}{i}{m}$=$\pf{P}{e}{m}$=34.4 \pm 0.1$Myr; this pattern is more easily seen after $300$ Myr, in the lower half of the black ensemble, rather than in the green orbit, where other patterns strongly affect the shape of its envelope. The second $e$ frequency, \pf{P}{e}{o}, is much more subtle, as it does not appear directly, but is connected to all of the new frequencies. Of the new frequencies present in $i$, we can see: a) there is a pattern that can be seen in the black points, made up of moments where the $i$ is allowed to deviate from the average inclination (\pf{P}{i}{p}$=1.8711\pm0.0010$ Myr); this pattern is similar to the allowed regions that give rise to \pf{P}{e}{p}, except with allowed regions at half the oscillation period instead of one-fourth of the period). b) The basic oscillation period of the green curve, \pf{P}{i}{o}, is slightly faster than twice the pattern period \pf{P}{i}{p} ($2\ppf{P}{i}{p}>$\pf{P}{i}{o}$=3.688 \pm 0.002$ Myr) as the $i$ oscillation is not locked with the allowed pattern in the same way it was on $e$; while it is not easy to disentangle all oscillations, this frequency has an amplitude of approximately $1.6^{\circ}$; and it has the right frequency and amplitude to show to be analogous to the main oscillations of $i$ in Eris and \dpn{2015}{KH}{162}, i.e. it originates from the regression of Pluto's line of nodes with respect to the Solar System's invariant plane. c) Finally, there is a slow modulation (\pf{P}{i}{sm}) of the amplitude with approximately 7.5 cycles in 1 Gyr; in the absence of DP perturbations this pattern is \pf{P}{i}{sm}$=133 \pm 5$ Myr \citep[see also][for a similar slow modulation]{Applegate86}.

The pattern with period \pf{P}{i}{p} is difficult to see at the beginning of the simulations, as it requires a background of different phases from the ensemble (i.e. it involves loss of coherence from the ensemble); however, starting a $\sim 175$ Myr this pattern remains stable until the end of the simulations; this pattern is formed by the points in time when the inclination is allowed to deviate the most from the average of $i \sim 15.7 ^{\circ}$. The true oscillation frequency at times seems to follow this pattern, at half the frequency; however, as mentioned before, it is slightly faster (3.742 vs. 3.688 Myr), and approximately every 133 Myr the green curve slips one notch or half a cycle:
\begin{equation}
{1\over \ppf{P}{i}{o}}-{1\over 2\ppf{P}{i}{p}}={1\over 266{\rm Myr}}={1\over2\ppf{P}{i}{sm}}.
\end{equation}
Also, whenever the green $i$ curve is fully in phase with the black pattern, there is an anti-correlation between $e$ and $i$ (e.g. 950 Myr, third column of figure \ref{fig:Pluto_evol}), while when the green curve is slipping from the pattern there is a correlation between $e$ and $i$ (e.g. 35 Myr and 575 Myr, first and second columns of figure \ref{fig:Pluto_evol}); overall every 133~Myr the evolution of $e$ and $i$ slip by one cycle: 
\begin{equation}
{1\over \ppf{P}{i}{o}}-{1\over \ppf{P}{e}{o}}={1\over 133{\rm Myr}}={1\over\ppf{P}{i}{sm}}.
\end{equation}
This also means that the difference in frequency between the oscillations and the pattern in $i$ is the same as the difference in frequency between the pattern in $i$ and the oscillations in $e$:
\begin{equation}
{1\over \ppf{P}{i}{o}}-{1\over 2\ppf{P}{i}{p}}={1\over 2\ppf{P}{i}{p}}-{1\over \ppf{P}{e}{p}}.
\end{equation}
Overall, each 133 Myr, there is exactly one more cycle in \pf{P}{i}{o} than in \pf{P}{e}{o}, with (2\pf{P}{i}{p}) lying exactly in the middle: 
\begin{eqnarray}    
\nonumber 133{\rm Myr}&=&\ppf{P}{i}{sm}\\
\nonumber &=&35.039\ppf{P}{e}{o}\\
\nonumber &=&35.539(2\ppf{P}{i}{p})\\
 &=&36.039\ppf{P}{i}{o}.
\end{eqnarray}

The importance of \pf{P}{i}{p}, as well as of \pf{P}{e}{p}, is not so much their values, but the existence of the patterns themselves, which only become apparent when an ensemble of orbits is used. These patterns show the existence of allowed and forbidden regions connected mainly to the evolution of Neptune, which in turn responds to the evolution of the planetary system as a whole, as we will see below.

In the bottom panels of Figures \ref{fig:Pluto} and \ref{fig:Pluto_evol}, we present the evolution of $q$ and $Q$ for Pluto. As in the previous cases, since the semimajor axis variations are small, the oscillations observed in both $q$ and $Q$ are dominated but the $e$ oscillations and share the same periods. Also of note is that these oscillations in $e$ are large enough to shift $q$ from below $a_{\it Nep}$ (as it is at present) to above $a_{\it Nep}$ i.e. from periods when Pluto's orbit crosses Neptune's orbit to periods when it remains outside it. Either way, there are no close encounters due to the detailed periods of the 3:2 resonance in which Pluto and Neptune lie.

\begin{figure*}[ht!]
\includegraphics[width=\linewidth]{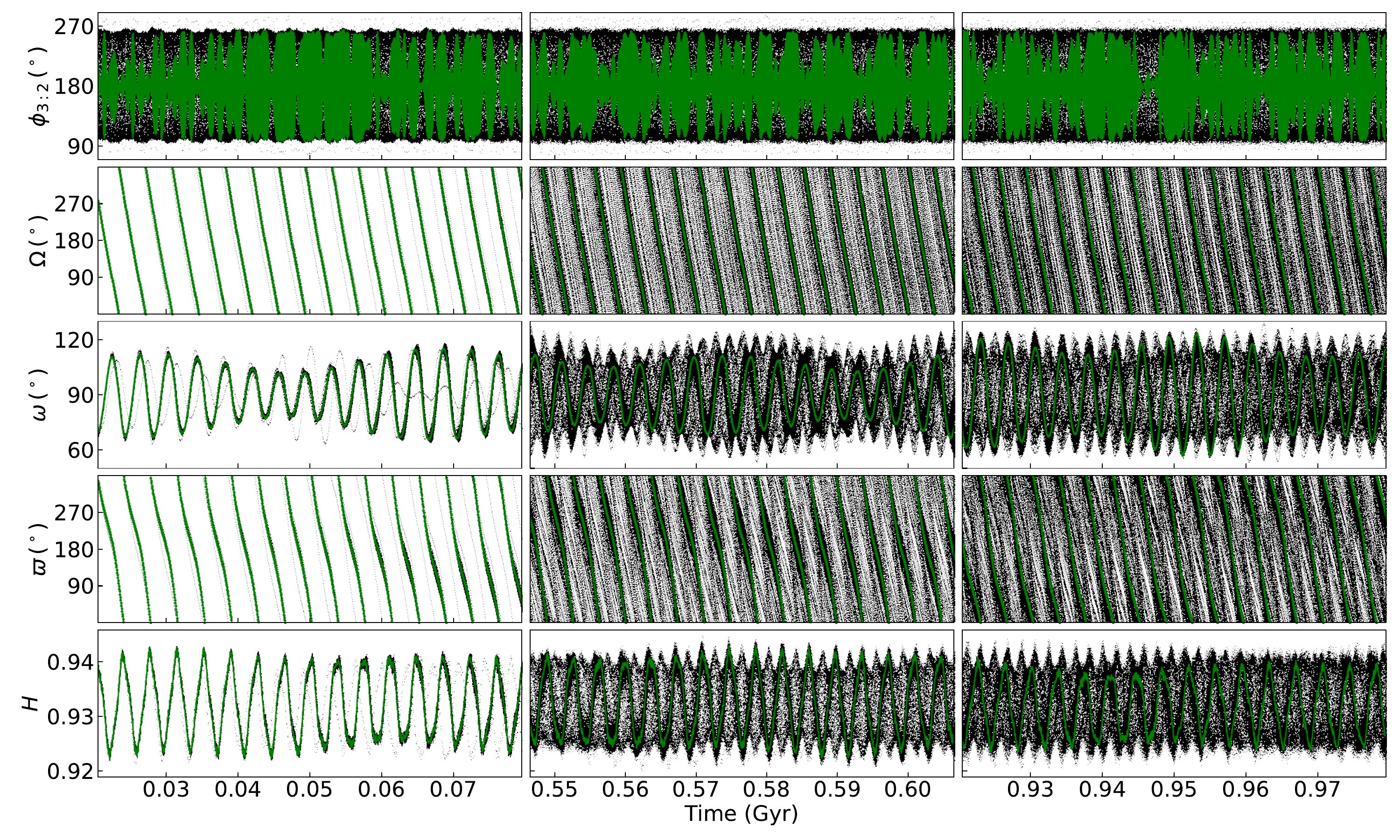}
\caption{Similar to figure \ref{fig:Eris_angles} but for Pluto's angular elements and Kozai's constant, as well as the resonant argument of the 3:2 MMR, $\phi_{3:2}$, presented in the first row of panels. Due to the rapid oscillations and circulations of these variables, we only show the three-time intervals presented in figure \ref{fig:Pluto_evol} for better comparison. \label{fig:Pluto_evol_angles}}
\end{figure*}

In Figure \ref{fig:Pluto_evol_angles}, we present the evolution of Pluto's resonant argument, $\phi_{3:2}=3\lambda_P-2\lambda_{Nep}-\varpi_P$, where $\lambda_P$ and $\varpi_P$ are the mean longitude and longitude of perihelion of Pluto, respectively, and $\lambda_{Nep}$ is the mean longitude of Neptune. We also present Pluto's angular parameters $\Omega$, $\omega$, $\varpi$, and $H$ as we did for Eris in Figure \ref{fig:Eris_angles}. However, as was the case for the main orbital parameters, the oscillations of Pluto's orbital angles evolve faster. We will only present three 60 Myr intervals, the same intervals that we presented in figure \ref{fig:Pluto_evol}. Again, the left column spans from 20 to 80 Myr, the middle column from 546.5 to 606.5 Myr, and the right column from 920 to 980 Myr.

The most notable feature in the upper panels of figure \ref{fig:Pluto_evol_angles} is the clear delimitation of the maximum excursions in the evolution of $\phi_{3:2}$, which quickly oscillates around the 180$^\circ$ mark, with the same frequency as $a$, and with an amplitude of $\approx 90^\circ$. Also, we note the anti-correlation, which can be traced with the green orbit, between the evolution of $a/a_0$ and $\phi_{3:2}$, this is to be expected as librating orbits trace ellipses in the $a$-$\phi_{3:2}$ plane. 

In the second row of panels of figure \ref{fig:Pluto_evol_angles}, we can see how $\Omega$ circulates with a nearly constant speed through each circulation and across all the orbits along the integration; the green orbit completes each cycle in $\ppf{P}{\Omega}{c}\approx3.688\pm0.002$ Myr, naturally, the same period of the main oscillation found in $i$, $\ppf{P}{i}{o}=3.688\pm0.002$ Myr; this value is also consistent with previous works, e.g. \citet{Applegate86,Milani89,Kinoshita96}. While this period is very regular for the 200 realizations, the phase is slightly perturbed when close encounters are present. Overall, the ensemble of orbits mostly maintains cohesion for about 150 Myr, and the number of revolutions for the ensemble can be identified for up to 400 Myr; after this mark, the faster and slower orbits start to overlap, eventually filling all the available range.

In the panels of the middle row of Figure \ref{fig:Pluto_evol_angles}, we present the evolution of $\omega$. Here, we can see how $\omega$ librates around $90^\circ$ with varying amplitude along the integration. The maximum amplitude values reach approximately $24^\circ$ in the first hundreds of Myr of integration, while near the end this increases to maximum amplitudes of $\approx33.5^\circ$. The evolution of $\omega$ presents 3 periods: the main oscillation has a period $\ppf{P}{\omega}{o}=3.80\pm0.01$, consistent with the oscillation in eccentricity $\ppf{P}{e}{o}=3.795\pm0.002$; an overall modulation is also present with $\ppf{P}{\omega}{m}=34.4\pm0.1{\rm Myr}=\ppf{P}{e}{m}=\ppf{P}{i}{m}$, with a closer similarity with the shape of the $e$ modulation since $\omega$ does not have the second modulation present in $i$; these values are again consistent with works from the literature, e.g. \citet{Milani89,Kinoshita96}. Finally, a pattern in the ensemble of orbits (black region) equal to the inclination pattern, $\ppf{P}{\omega}{p}=\ppf{P}{i}{p}=1.8711\pm0.0010$ Myr; this pattern and its period are not usually seen, since previous works considered only a small number of orbits and not a large ensemble, which is required for this pattern to emerge. Again the slight difference between \pf{P}{\omega}{o} and $2\times\ppf{P}{\omega}{p}$ produces an additional modulation on the amplitude of the individual curve's oscillations. This can be seen in the green curve in all three panels and the black region for the center and right panels.

As identified in previous works, the 34.4 Myr $\omega$ modulation, also present in $e$ and $i$, denotes the correlation of these variables through Kozai's constant. This modulation has been identified as a secondary resonance from the vZLK mechanism, which originates from the commensurability of the libration period of $\omega$ and the circulation period of $\Omega-\Omega_N$, where $\Omega_N$ is Neptune's longitude of the ascending node. This resonance was called ''super resonance'' by \citet{Milani89} and is the narrowest of a triplet of resonances that Pluto can simultaneously lie in \citep{Wan01}. It should be noted that $\Omega$ does not present such a period, since it is related to $i$ independently of the vZLK resonance.

In the panels of the fourth row of figure \ref{fig:Pluto_evol_angles}, we present the evolution of $\varpi$. Being the sum of $\Omega$ and $ \omega$ it circulates with a period of $\ppf{P}{\varpi}{c} = 3.688\pm 0.002$ Myr, but it does not maintain the same speed across the entire cycle; we can see a small hump at each cycle of the green curve; even more, with $\Omega$ and $\omega$ having slightly different frequencies, the hump slowly shifts to lower phases at each circulation with $\ppf{P}{\varpi}{sc}= 143\pm 10$ Myr, which is consistent with the slow modulation present on $e$ and $i$, $\ppf{P}{\varpi}{sc}\approx\ppf{P}{e}{sm}=\ppf{P}{i}{sm}$. Finally, the slow modulation of $\omega$ (with period $\ppf{P}{\omega}{m}=34.4$ Myr is shown in the black points as steps in the black shadows that accompany the (-more regular-) green hump, producing a pattern with a period of $\ppf{P}{\varpi}{m} = 27.7\pm0.1$ Myr, where $1/\ppf{P}{\varpi}{m}=1/\ppf{P}{\varpi}{sc}+1/\ppf{P}{e}{m}$.

It is interesting to compare Pluto's three orbital angles ($\Omega$, $\omega$, and $\varpi$) with those of Eris; we find that, for Eris, $\varpi$ is librating, in fact it is nearly constant, i.e. Eris's perihelion always points in the same direction; Pluto instead has a librating $\omega$ showing how its perihelion is always located above the invariant plane.

In the last row of panels of Figure \ref{fig:Pluto_evol_angles}, we present the evolution of Kozai's constant; while for Eris $H$ was a simpler curve than $e$ or $i$, for Pluto $H$ acquires the sum of the characteristics of both $e$ and $i$. Its main oscillation (seen more clearly in the green curve) has a period $\ppf{P}{H}{o}=3.688\pm0.002$ Myr, that corresponds to that of $i$ and $\Omega$, $\ppf{P}{i}{o}=\ppf{P}{H}{o}=\ppf{P}{\Omega}{o}$. While the main period of $H$ and $\Omega$ are equal, as is the case in Eris, in Pluto's case $H$ has other smaller but non-negligible components; from a Fourier transform, we can find several harmonics of \pf{P}{\Omega}{c}, as well as components with periods of \pf{P}{e}{o} and its harmonics.

The small difference between \pf{P}{e}{o} and \pf{P}{i}{o} will create a slow modulation in the amplitude of $H$, again with a frequency $1/\ppf{P}{H}{sm}=1/\ppf{P}{i}{sm}=1/ \ppf{P}{i}{o} - 1/\ppf{P}{e}{o}$; this amplitude is slightly larger when $e$ and $i$ are anti-correlated than when they are correlated. 

The existence of permitted and forbidden regions, in both $e$ and $i$, creates analogous regions in the black area formed by the 200 $H$ realizations. Overall the notches present in $i$'s permitted region are intrinsically larger and better defined than those of $e$; when they combine, there appears a unique interplay in the permitted regions of $H$: when there is correlation or anticorrelation, the notches present in $i$ seem to dominate the behavior, as a matter of fact, those of $e$, do not seem to be present, however when $e$ and $i$ are $90^\circ$ out of phase, these notches seem to disappear and a faster pattern (similar to the one of $e$) seems to appear (e.g. at 965 Myr or 550 Myr); upon closer inspections, it is clear that both patterns are present, as the notches at those times are seen more clearly on the top end than at the lower end of the range, and the notches are not uniformly spaced. 

Finally, the modulation period, \pf{P}{H}{m}=34.4 Myr, appears subtly here: when the amplitude of the oscillations of $e$ or $\omega$ is bigger, the oscillation amplitude seems to be slightly larger on average while, when those amplitudes are small, $H$'s central value seems to be slightly lower ($\sim0.9327$ vs. $\sim0.9315$); this effect (with a $\Delta H \approx 0.001$) is much smaller than the full width of the black region ($\Delta H \approx 0.020$), and even than that of the ``narrowest allowed regions'' ($\Delta H \approx 0.015$) seen in black once most of the realizations have lost coherence after 250 Myr.

\begin{figure}[ht!]
\includegraphics[width=\linewidth]{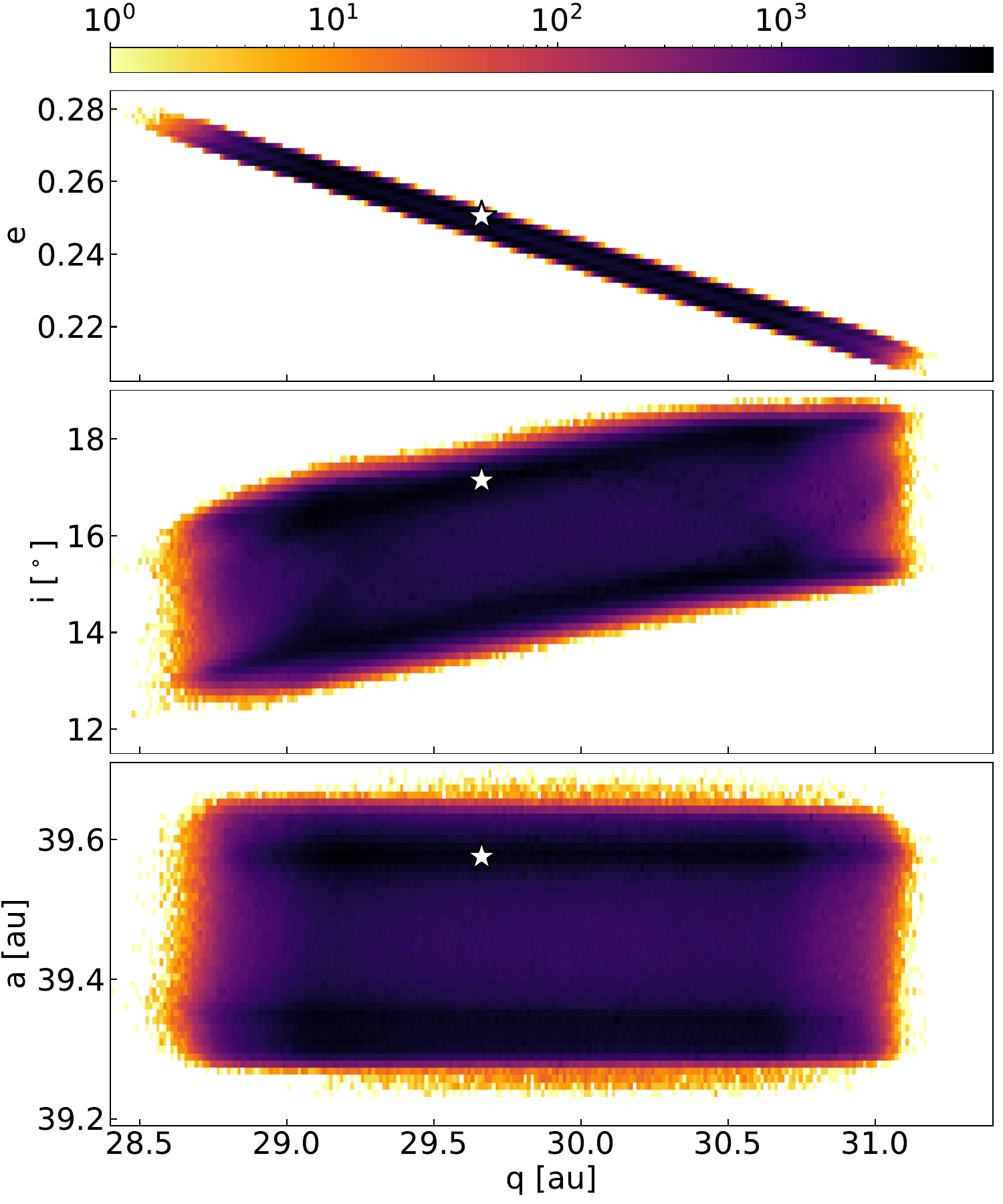}
\caption{Same as Fig. \ref{fig:Eris_density} but for Pluto`s orbital density. \label{fig:Pluto_density}}
\end{figure}

Figure \ref{fig:Pluto_density} is equivalent to figures \ref{fig:Eris_density} and \ref{fig:2015KH162_density}, but for Pluto. As we did for \dpn{2015}{KH}{162}, all axes are presented linearly; this was done since the range in all cases is too small to warrant using logarithmic scales.

The short range of the figure shows that all of Pluto's realizations are well-behaved, and none drifts too far away from its current values. This is similar to the \dpn{2015}{KH}{162} case, but the origin of their behaviors is quite different. The stability of \dpn{2015}{KH}{162} is easily understandable due to its remoteness from any strong perturbation, while in the case of Pluto, the small phase-space volume occupied by the 200 realizations results from the constraining effect of the 3:2 MMR, which restores Pluto's orbital parameters to their preferred values after each perturbation.

From figure \ref{fig:Pluto_density} we can see the small variations in $a$, $\Delta a\approx0.45$ au, which are $\sim6$ times smaller than the allowed variations in $q$, $\Delta q\approx2.8$ au. Such small $a$ variations force a strong anti-correlation between $e$ and $q$, with $\Delta e\approx0.07$, and a correlation with high dispersion between $i$ and $q$, with $\Delta i\approx7^\circ$. The black bands shown in the upper and lower regions of the $a-q$ and $a-i$ panels are due to the sinusoidal nature of these oscillations; the fact that these regions are poorly defined towards the edges is due to the variability of the center and the amplitude of such oscillations (as can be seen in figures \ref{fig:Pluto}, \ref{fig:Pluto_evol}, and \ref{fig:Pluto_evol_angles}. Finally, we note that the ``orange halo'' around the $a-q$ and $i-q$ planes, as well as the extremes in the $e$ panel, is due to a single realization, that is slightly less constrained than the other 199, it can also be seen in the evolution of $a$ and $\phi_{3:2}$ (fist panel of Figure \ref{fig:Pluto_evol_angles}). 

%\subsubsection{Relation with the fundamental secular frequencies of the GPs}

\subsubsection{Pluto's pattern periods relation with the fundamental secular frequencies of Neptune}

For an object at Pluto's distance from the Sun, most periods discussed in section 4.1 represent the natural evolution of the orbital parameters. In general, we expect that perturbations from other DPs would result in small phase shifts, after which the natural periods would resume while retaining the phase shift. However, close encounters do not seem to affect the phase of 
%some patterns presented before: specifically, they do not seem to affect 
the patterns associated with \pf{P}{i}{p}, \pf{P}{e}{p}, and \pf{P}{\omega}{p}; instead those patterns are a reflection of how the main dynamical characteristics of the system (i.e. Sun+GPs) restrict the liberties of Pluto within Neptune's 3:2 MMR.

In particular, from Neptune's secular frequencies, associated with the longitude of pericenter and longitude of the ascending node, $g_8$ and $s_8$ respectively \citep[e.g.][]{Carpino87,Laskar88,Knezevic91}, we find that $\ppf{P}{i}{p} = \ppf{P}{\omega}{p} = \left|1/s_8\right|$ and it seems likely that \pf{P}{e}{p} comes from $1/\ppf{P}{e}{p}= g_8 - s_8$. 

The above result means that the pattern component observed in the $i$ ensemble oscillations does not come from $e$ or the vZKL mechanism, but rather from the precession of Neptune's orbit around the invariable plane, hence \pf{P}{i}{p} is directly linked to  $\left|1/s_8\right|$, Neptune's fundamental secular frequency of the longitude of ascending node. But it also means that Pluto's orbit has to be in phase with Neptune in one of these two allowed phases. The maxima of the pattern oscillations of $e$ are also related to Neptune's orbit; in this case, they occur only when Neptune's fundamental frequencies of the longitude of pericenter and the longitude of ascending node synchronize with each other, hence $1/\ppf{P}{e}{p}$ comes from the difference of $g_8$ and $s_8$. This is a frequency that has not been evident without the advantage of having an ensemble of Plutos, thus it was not found in previous studies of the outer planets and Pluto \citep[e.g.][]{Milani89,Malhotra97}.

In the Solar System, Pluto is constantly perturbed from its idealized orbit by close encounters with other DPs. Yet these encounters will only be able to perturb Pluto for a brief time. In the end, the locking into resonance with Neptune will force Pluto back into one of the allowed phases determined by these 2 fundamental patterns.

%However, other combinations between fundamental secular frequencies of the GPs \citep[e.g.][]{Nobili89} may be responsible for the different periods that we see in Pluto, it is not clear how significant such combinations could be to the overall evolution of Pluto, due to the additional small perturbations in our model introduced by the presence of the other DPs. A deeper analysis of these frequencies' significance deserves a careful study but lies beyond the scope of this paper. 

\subsection{\dpn{2010}{EK}{139}}\label{sec:EK139}

%\begin{figure*}[ht!]
%\includegraphics[width=\linewidth]{2010EK139_evol_green.pdf}
%\caption{2010EK...  \label{fig:2010EK13_evol_2}}
%\end{figure*}

\dpn{2010}{EK}{139}, recently named ``Dziewanna'', is one of the brightest TNOs \citep{Pal12}. Regarding \dpn{2010}{EK}{139}'s magnitude, it should be noted that its current absolute magnitude is 4.09, slightly larger than our original threshold of $H_V=4$, used in \citetalias{Munoz21}; however since the object is already part of our simulations we have kept it for our analysis.

%By 2019, the absolute magnitude of this object listed in the MPC was 3.90, which placed it within our selection criteria for the most massive objects used in \citet[][where we used a cutoff of $H_V=4$ on the value of the absolute magnitude listed in the MPC]{Munoz19}. Today's listed absolute magnitude of 201EK is 4.09\footnote{\url{}}, putting it slightly below the threshold used in \citetalias{Munoz21}. However, the original cutoff is completely arbitrary, and the point of selecting the most massive objects for a given simulation is to use the most relevant objects for a finite amount of CPU power, however we already have 200 high quality realizations of \dpn{2010}{EK}{139}, and such downgrade does not really diminish the quality of the evolution of the other objects.

\dpn{2010}{EK}{139} lies within Neptune's 7:2 MMR; this resonance is much weaker than the 3:2 MMR. Due to the resonance inherent strength, the specifics of \dpn{2010}{EK}{139}`s orbital parameters, and close encounters with other minor bodies, in general, the presence of \dpn{2010}{EK}{139} within the MMR is intermittent. However, for the next 60 Myr, it will remain locked within the resonance; also, while most realizations leave the resonance at least once, 20 remain within the resonance for the entire 1~Gyr integration. On the other hand, while most integrations drift in and out of the resonance, some leave it and do not return to the resonance by the end of our simulations.

%Las realizaciones que encontramos resonantes son: 21, 31, 36, 42, 46, 48*, 61, 65, 71, 84, 92, 133, 135, 139, 142, 150, 160, 166, 176*, 179* Nota en particular a las 3 con asterisco, pues tienen picos altos, aunque definitivamente no llegan al umbral de la circulacion

\begin{figure*}[ht!]
\includegraphics[width=\linewidth]{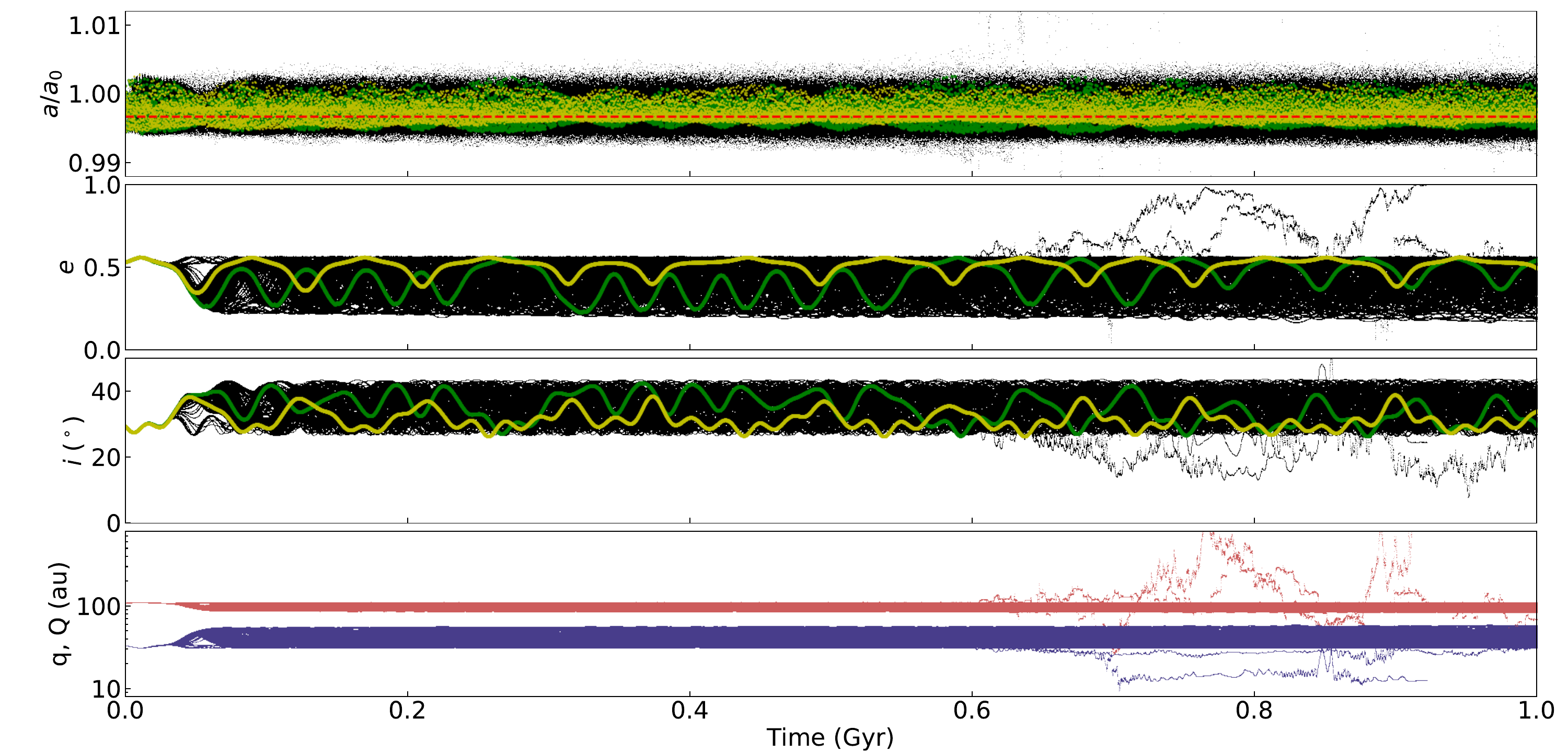}
\caption{Evolution of the main orbital parameters of \dpn{2010}{EK}{139}'s 200 realizations. Similar to Fig. \ref{fig:Eris}, but with two highlighted realizations, in green and yellow for $a$, $e$, and $i$, to exemplify the variety of behaviors. The dashed red line in the top panel represents the nominal location of the 7:2 MMR. 
\label{fig:2010EK13_evol}}
\end{figure*}

In Figure \ref{fig:2010EK13_evol}, we present the evolution of \dpn{2010}{EK}{139} $a$, $e$, $i$, $q$, and $Q$, similarly as we did for Eris, \dpn{2015}{KH}{162}, and Pluto in Figs. \ref{fig:Eris}, \ref{fig:2015KH16}, and \ref{fig:Pluto}, respectively. The main difference with the previous figures is the presence of two representative orbits; we do this because, in contrast to the previous objects, the behavior of \dpn{2010}{EK}{139}`s realizations is not uniform; we thus highlight (in green and yellow) more than one realization to represent the richness of its orbital behavior better.

The first panel of Fig. \ref{fig:2010EK13_evol} shows the evolution of $a/a_0$ for \dpn{2010}{EK}{139}, where $a_0=69.62089$ au. Out of the 200 realizations, 198 formed a band that thickens with time, in contrast to the resonant band formed by Pluto's realizations; this exemplifies the weaker nature of the 7:2 MMR. The other two realizations are destabilized and move far from the resonance, with one getting ejected from the Solar System. Over the black band we highlight two different orbits, the green dots show a realization with an intermittent resonant behavior, while the yellow one shows a realization with an orbit that remains well-locked within the resonance for the whole 1 Gyr. The red dashed line shows the nominal location of the 7:2 MMR, $a_{7:2}/a_0=0.99677$, where we take again $a_{7:2}=\left<{a_{Nep}}\right>(7/2)^{2/3}$ with $\left<{a_{Nep}}\right>=30.10411$ au. Overall, the black band is not symmetric around the nominal resonance location. However, the average $a$ is only slightly larger than $a_{7:2}$; equivalently, the average of $(a/a_{7:2})^{(-3/2)}$, the angular orbital speed compared to the resonance, is slightly smaller than perfect resonance. These values will result in an average of 10 circulations per realization. Later on, when studying $\phi_{7:2}$, we will show that all the circulations of \dpn{2010}{EK}{139} are regressing (i.e. fall behind the resonance). 

Regarding our two selected realizations, we can see how both orbits pop above and below the resonance location; this is expected for orbits trapped or nearly trapped in resonance; to be resonant, the local average of such points has to coincide with the resonance nominal location, for however long the object is locked within the resonance. In both cases we can see many more points with $a<a_{7:2}$ than $a>a_{7:2}$, yet, the points with large semimajor axis deviate more from $a_{7:2}$ thus keeping the overall average. Upon closer examination, the dispersion of the green dots varies with time, giving an appearance of elliptical beads of different lengths in a string; when the scatter diminishes, the lower limit gets very close to the resonance thus disturbing the balance that is achieved everywhere else in the green dots (as well as in the yellow dots); in these moments the average shifts upward from $a_{7:2}$. This happens 11 times for the green orbit; at these times \dpn{2010}{EK}{139}'s resonant argument shows a regressive circulation.

The second panel of Fig. \ref{fig:2010EK13_evol} shows the evolution of $e$ for \dpn{2010}{EK}{139}. During the first 30 Myr, there is virtually no variation for the 200 realizations, at that point the bundle of realizations splits in two, and over the next 30 Myr, it loses coherence so that at 60 Myr there is a dark band with $0.22 < e < 0.57$; this band will slightly broaden to $0.19 \lesssim e < 0.57$ by the end of the simulation. At approximately 600 and 625 Myr, we can see the two realizations that become destabilized and move far from the resonance region. Over the black band, we have plotted our two representative realizations, in green and yellow; these curves present non-periodic oscillations, with the green realization presenting 17 oscillations and the yellow one only 10. When looking at all realizations it becomes clear that the number, the height, and the width of these oscillations behave chaotically, within the allowed black band.

From the yellow realization, we can see that all its oscillations lie slightly high on the black band with $e>0.33$ at all times, while 13 of the green oscillations go much lower. When looking into all 200 realizations, for the first 200 Myr there is a one-to-one correlation between circulations of $\phi_{7:2}$ and oscillations of $e$, where $e_{min}<0.299$; it should be noted that for larger timescales this boundary becomes slightly fuzzy, but the overall trend where circulations of $\phi_{7:2}$ only occur for oscillations where $e_{min}\lesssim0.30$ remains. This result lines up with the expected shape and width of MMRs \citep[i.e. a broadening of the resonance with increasing $e$, e.g.][]{Murray99}. Note that \citet{Malhotra18} find that for the resonances 3:2, 2:1, and 5:2 there is a maximum resonance width at mid-eccentricity values, a trend we expect to remain for the 7:2 MMR.

The third panel of Fig. \ref{fig:2010EK13_evol} shows the evolution of $i$ for \dpn{2010}{EK}{139}. As in our previous objects, due to vZLK's mechanism, $i$ responds to every oscillation in $e$ with an opposite oscillation at the same point in time; also, as in all the previous objects, there is an additional component to these $i$ oscillations. In the case of Eris and \dpn{2015}{KH}{162}, we saw that the component originating from $e$ is much smaller than the new component (absent in $e$ and appearing in $H$ and $\Omega$). For Pluto, both components were of similar size. For \dpn{2010}{EK}{139} we see that the new component is about 5 times smaller than the $e$ one; this new component again corresponds with the regression of \dpn{2010}{EK}{139}'s line of nodes with respect to the Solar System's invariant plane, but due to its small amplitude relative to the $e$ contribution, this oscillation does not represent the main component of $i$'s oscillations for this object, unlike the other objects presented in this work. Also, unlike in the previous objects, this component behaves chaotically (as did the oscillations in $e$). This new component is slower to lose coherence; at 60 Myr the ensemble of $i$ behaves as an oscillating band, that will require more or less 140 Myr to lose coherence and fill the entire $26.5^{\circ}<i<43^{\circ}$ range; again, this will spread a little by the end of the simulation with $25.5^{\circ}<i<44^{\circ}$. We can also see the two realizations that escape the bundle at 600 and 625 Myr. Finally, the overplotted green and yellow lines show the chaotic behavior of this variable.

Finally, the bottom panel of Fig. \ref{fig:2010EK13_evol} shows the evolution of $q$ and $Q$ for \dpn{2010}{EK}{139}. The most notable feature in this panel is showing how the two destabilized orbits move far away from the Sun (without immediately escaping the Solar System); at the same time, the $q$ of these two realizations dips inside the orbit of Neptune, with one of them going as far in as reaching Saturn, which will have drastic effects, ejecting such realization from the Solar System.

\begin{figure*}[ht!]
\includegraphics[width=\linewidth]{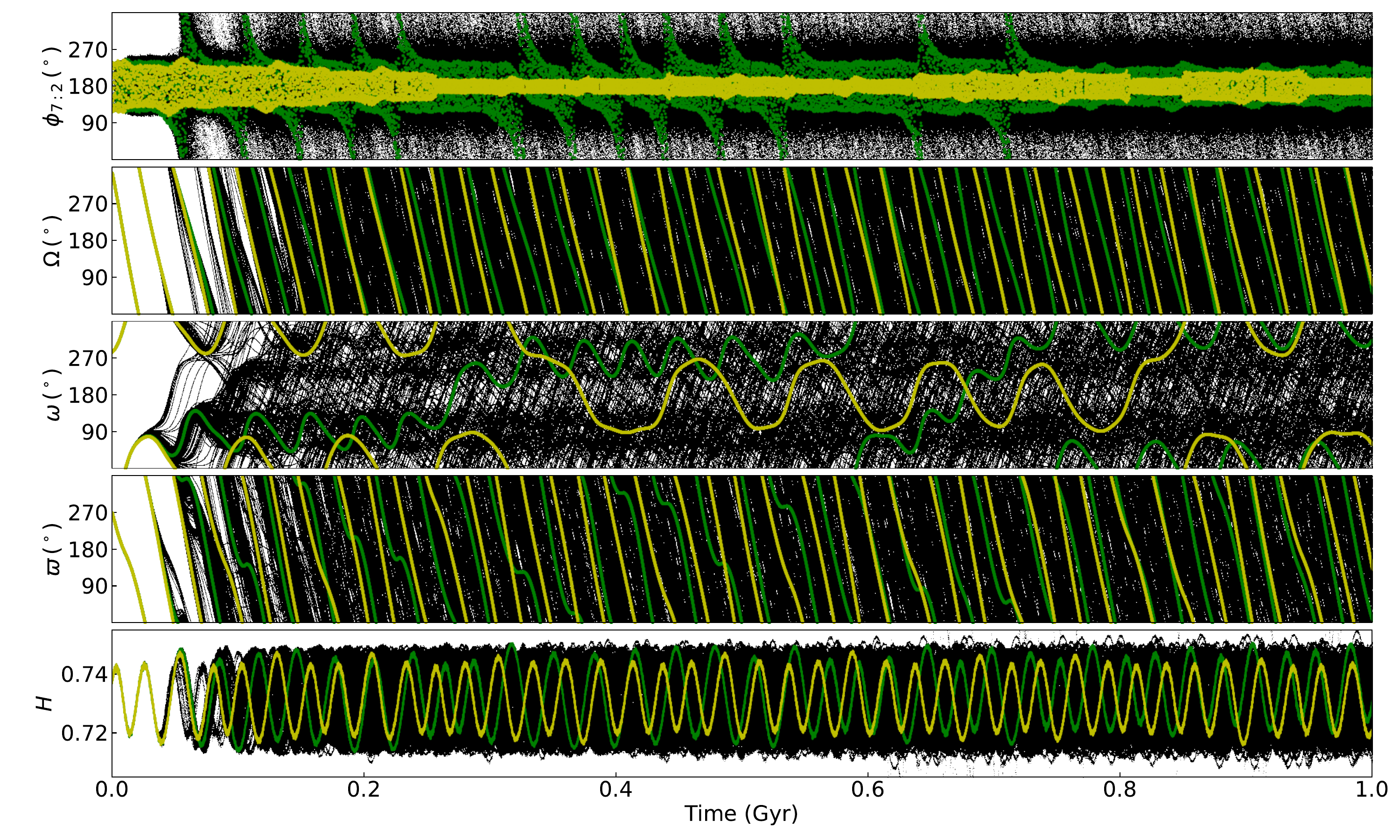}
\caption{Similar to figure \ref{fig:Eris_angles} but for \dpn{2010}{EK}{139}'s angular elements and Kozai's constant, as well as the resonant argument of the 7:2 MMR, $\phi_{7:2}$, presented in the top panel. As in Fig. \ref{fig:2010EK13_evol} we highlight two orbits, in green and yellow, to exemplify the variety of orbital behaviors. \label{fig:2010EK13_angles}}
\end{figure*}

\begin{figure}[ht!]
\includegraphics[width=\linewidth]{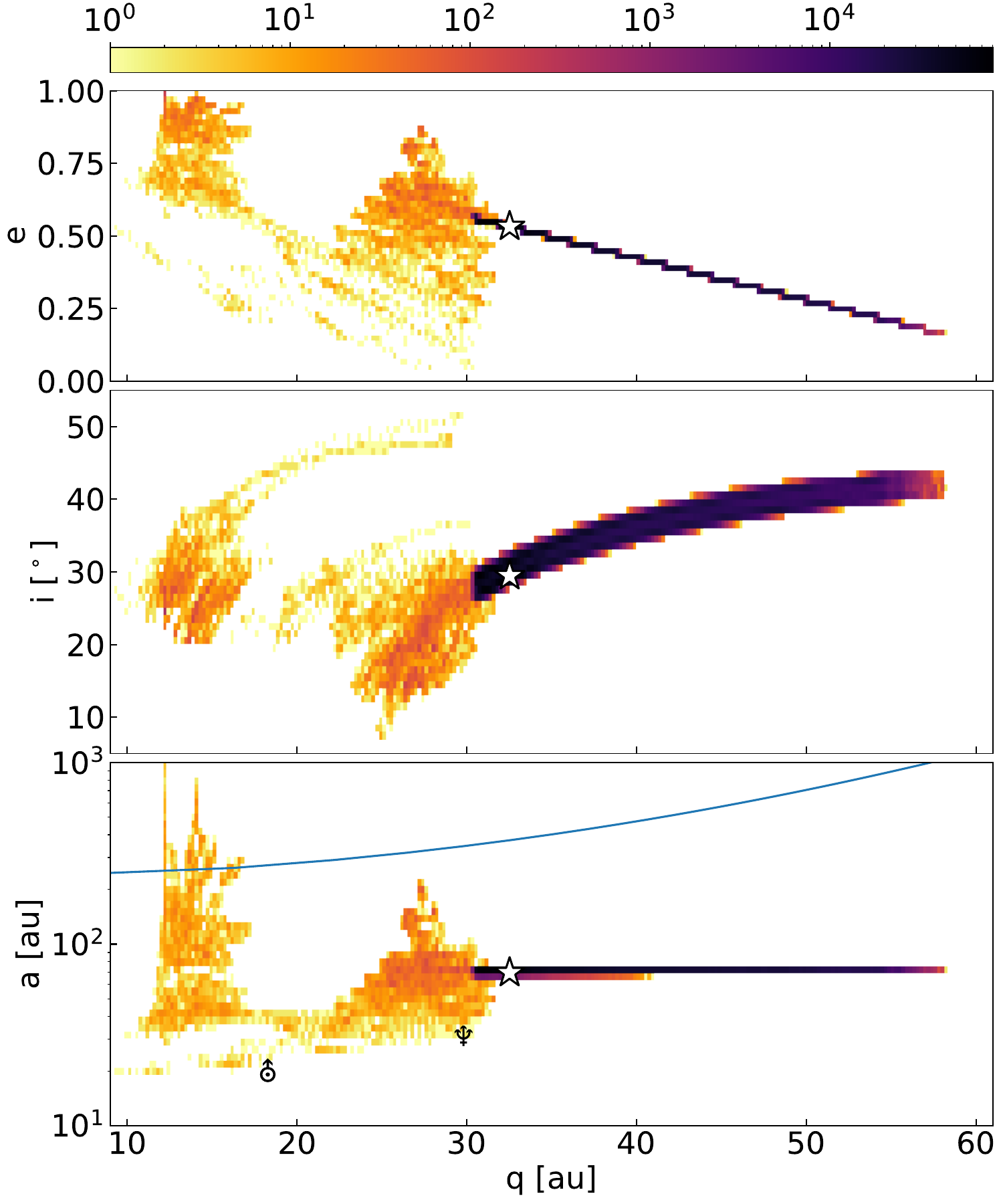}
\caption{Same as Fig. \ref{fig:Eris_density} but for \dpn{2010}{EK}{139}'s orbital density. In the bottom panel, the blue line shows the stability limit for large $a$ objects determined by \citet{Batygin21}; the symbols show the location of Neptune and Uranus in the $q$ v.s. $a$ plane. \label{fig:2010EK13_density}}
\end{figure}

In Figure \ref{fig:2010EK13_angles}, we present the evolution of \dpn{2010}{EK}{139} $\phi_{7:2}$, $\Omega$, $\omega$, $\varpi$, and $H$, similarly as we did for Eris and Pluto in Figs. \ref{fig:Eris_angles} and \ref{fig:Pluto_evol_angles}, respectively. For \dpn{2010}{EK}{139} we study the libration angle given by $\phi_{7:2}=7\lambda_D-2\lambda_{Nep}-5\varpi_D$, where $\lambda_D$ and $\varpi_D$ are the mean longitude and the longitude of perihelion of \dpn{2010}{EK}{139}, respectively, and $\lambda_{Nep}$ is the mean longitude of Neptune. We include the same two representative orbits highlighted in green and yellow, as in figure \ref{fig:2010EK13_evol}.

The top panel of Fig. \ref{fig:2010EK13_angles} shows the evolution of $\phi_{7:2}$ for \dpn{2010}{EK}{139}. Placing black dots for our 200 realizations forms the band that librates around 180$^\circ$ for the first 60 Myr after which two things happen: there start to be excursions out of the core of the band, showing the constant presence of circulation within the ensemble; and an overall broadening of the core of the band with time, showing how random close encounters with other DPs, alter the level at which \dpn{2010}{EK}{139} is trapped within the resonance, with the width of the band dictated by the realizations with the largest librations. 

The yellow dots show an example of a realization trapped within the 7:2 MMR during the entire Gyr; recalling that 20 out of our 200 realizations never circulate, the yellow realization has relatively tight librations. From the fully resonant orbits, some have librations of more than $180^{\circ}$ for most of the integration; also, several of them have oscillations of $e$ that bring them close to circulation limit that can widen the librations close to $270^{\circ}$, reaching angles close to $0^{\circ}$ and to $360^{\circ}$ at different periods of the ``disturbance''. On the other hand, the green dots show an orbit that circulates 11 times. These circulations coincide with the 11 large dips in $e$; the four smaller dips produce four sets of asymmetrical nubs after 750 Myr. When $e$ dips to values close, but larger than, $e=0.299$ those nubs extend forming spikes that can, in extreme cases, extend close to $0^{\circ}$ and to $360^{\circ}$; of course, if $e$ dips below the threshold, circulation will occur. There is a third kind of orbit, that is not shown in figures \ref{fig:2010EK13_evol} and \ref{fig:2010EK13_angles}, when the overall $e$ decreases enough so that not only the valleys but also the crests of the oscillations dip below the circulation threshold there will be very frequent circulations, with one realization reaching 70 circulations during the full integration; these orbits, while outside, lie very close to the 7:2 MMR and are affected by it. Finally, once the two unstable orbits move far from the resonance, $\phi_{7:2}$ becomes meaningless. 

%NOTA para $\phi$: when oscillations in eccentricity dip close to $e\gtrsim$0.299, the librating band of $\phi$ widens without quite reaching the circulation limit....... While oscillations in $e$ will produce asymetric deviations from the central band (zig-zags for large $e_{min}$, circulations for low $e_{min}$) the width of the librating band does not seem to depend much on the $e$ value.

The second panel of Fig. \ref{fig:2010EK13_angles} shows the evolution of $\Omega$ for \dpn{2010}{EK}{139}. $\Omega$ regresses constantly, as has happened for all previous objects; however, as mentioned when studying $i$, this circulation behaves chaotically with a wide range of periods varying between realizations, and across each realization; for \dpn{2010}{EK}{139} the typical circulation period is $25 - 28$ Myr, however, among all the ensemble, there are some cycles faster than 20 Myr and some slower than 35 Myr.

The third panel of Fig. \ref{fig:2010EK13_angles} shows the evolution of $\omega$. The two selected realizations mainly librate; however, in the full ensemble, broadly speaking there is a mixture of realizations where $\omega$ circulates, librates, or is a hybrid of the two behaviors; overall the most common behavior is upward circulation (precession), followed closely by libration, with a relatively small fraction of the realizations circulating downward (regressing). Libration can settle around $0^{\circ}$, $90^{\circ}$, $180^{\circ}$, or $270^{\circ}$, as can be seen, both in the highlighted realizations and, partially, in the integrated black shape; usually when libration occurs it remains stable for a few million years and then shifts to another quadrant. When considering all the realizations most of the libration occurs around $90^{\circ}$, followed by $270^{\circ}$, $0^{\circ}$, with $180^{\circ}$ having the least librations.

Strictly speaking, most of the time \dpn{2010}{EK}{139} is not in the vZLK resonance, but for some time intervals, the conditions for the vZLK mechanism are fulfilled, i.e. $\omega$ librating around 90$^{\circ}$ or 270$^{\circ}$. Circulation on $\phi_{7:2}$ occurs more frequently when omega crosses the $\omega=90^{\circ}$ or $270^{\circ}$ thresholds while going up; this is a required but not sufficient condition. This means that circulation on $\phi_{7:2}$ is more likely if $\omega$ is librating around $90^{\circ}$ or $270^{\circ}$ or if it is circulating upwards (either of these conditions will usually result in the circulation of $\phi_{7:2}$, although for neither condition it is guaranteed). On the other hand, libration of $\omega$ around $0^{\circ}$ or $180^{\circ}$, as well as downward circulation of $\omega$ guarantee libration of $\phi_{7:2}$, i.e. having \dpn{2010}{EK}{139} trapped in the 7:2 MMR. These arguments imply that this object is usually very unlikely to be trapped in MMR and experience the vZLK resonance simultaneously, unlike Pluto. For the 200 \dpn{2010}{EK}{139}'s realizations, simultaneous librations of $\phi_{7:2}$ (around 180$^{\circ}$) and $\omega$ (either around 90$^{\circ}$ or 270$^{\circ}$) only occur for approximately 2\% of the time; although there are several realizations where it happens for several hundred Myr, with the longest being slightly over 600 Myr. Due to these less stringent conditions, \dpn{2010}{EK}{139} is in general less stable than Pluto, such that even one of its realizations ends up being ejected from the Solar System in just 1 Gyr.  

%NOTAS EXTRA: loa libamientos de $\omega$ en $0^{\circ}$ or $180^{\circ}$ siempre son amplios $(\sim 160^{\circ}, 170^{\circ})$ y tardados $\sim 100$ Myr (excepto cuando es un solo doblez) en cambio los libramientos alrededor de $0^{\circ}$ or $180^{\circ}$ aunque tienen mayor dispersion de tamaños tienden a ser mas pequeños $(\sim 30^{\circ}, 120^{\circ})$ y rapidos $\sim 15 - 70$ Myr....

The fourth panel of Fig. \ref{fig:2010EK13_angles} shows the evolution of $\varpi$ for \dpn{2010}{EK}{139}. Being the sum of $\Omega$ and $\omega$, overall its behavior is dominated by the circulation of $\Omega$; thus $\varpi$ shows mostly circulation, with some occasional kinks produced by rapid oscillations in $\omega$. 

The last panel of Fig. \ref{fig:2010EK13_angles} shows the evolution of $H$ for \dpn{2010}{EK}{139}. $H$ shows oscillating patterns with variable periods and amplitudes, e.g. when looking into the yellow and green realizations, we find that the yellow one has, on average, a slightly smaller amplitude and is slightly faster than the green one; this is very different from the behavior for Pluto's and Eris's, where the periods were identical or nearly identical at all times for all realizations. The oscillations of the 200 realizations start synchronized, and again, they start losing coherence after 30 Myr, resulting in a dense band after just $\sim100$ Myr. There is, again, a correspondence between the period found in $H$ and the circulation period of $\Omega$; qualitatively, the secondary components of \dpn{2010}{EK}{139} are larger than those of Eris, which are almost non-existent, and much smaller than those of Pluto. On the other hand, the main component is again a sinusoidal wave linked to the phase of \dpn{2010}{EK}{139}'s ascending node, with respect to the ascending node of the Solar System's invariant plane; however, it does not follow the behavior of equation \ref{eq:H_Eris}, instead, the amplitude of the oscillations is more or less proportional to the circulation time of $\Omega$. Quantitatively we find that, for \dpn{2010}{EK}{139}, $H$'s slope correlates very strongly with:
\begin{equation}
\label{eq:H_EK139}
\frac{dH}{dt} \appropto \sin(\Omega-\Omega_{\it SSIP}), 
\end{equation}
where $\Omega_{\it SSIP}=107.58^\circ$ is the longitude of the ascending node of the invariable plane of the Solar System. The connection between slower circulations and larger amplitudes is through $e$. Despite having nearly identical $a$ and identical $\Delta i$, when $e$ is relatively small, the regression of $\Omega$ becomes less efficient, thus slower; on the other hand, in equation \ref{eq:H}, the $\sqrt{1-e^2}$ component becomes larger, and thus the amplitude of the oscillations, modulated by $i$ through the $\cos(i)$ component, becomes larger.

Finally, in figure \ref{fig:2010EK13_density}
we show the density plot of \dpn{2010}{EK}{139}'s realizations. As we observed in Eris, the large eccentricity variations with an almost constant semimajor axis, result in large excursions of \dpn{2010}{EK}{139}'s perihelion of almost 30 au. In the three panels, the purple/black area represents all 198 regular orbits, and a significant fraction of each of the two irregular orbits. The orange and yellow colors represent the irregular part of these last two orbits, where the right ``cloud'' with $21<q<32$ au represents the realization that remains bound to the Sun, and the left cloud mostly at $11<q<23$ au, but extending up to $q\sim30$ au at high $i$, represents the orbit that is ejected from the Solar System. With the blue line, we show the instability limit for large $a$ particles \citep[as found by][their equation 15]{Batygin21}. Our escaping particle is greatly affected the first five times it crosses this line and is effectively ejected in its sixth crossing, in contrast to the irregular orbit that remains in the Solar System, i.e. without crossing the instability limit.

\section{Conclusions
\label{sec:conclusions}}

In this work, we used results from 200, 1~Gyr long simulations of the outer Solar System, which consider the gravitational interaction of the Sun, the four GPs, and the 34 largest TNOs \citep[a.k.a. DPs, see][]{Munoz19}. Though the initial conditions of the 39 massive objects in our simulations are identical, there are small differences in the time steps taken by each realization \citepalias[for more details, see Section \ref{sec:N-body}, as well as][]{Munoz21}. Such differences in time steps lead to widely different evolutions according to the global stability of each object. These small deviations within the integrations should not be thought of as deficiencies, as the real Solar System contains thousands of smaller objects within the trans-Neptuninan region; these real objects are not massless, and their presence will perturb the evolution of our 34 TNOs in small and seemingly random ways, just as the deviations introduced by our integrators.

%realizations with different initial time steps of the simulation, while others result from the self-adaptive time-step produced by the Bulirsch-Stoer integrator, when it solves for close encounters between different collections of massless particles used to simulate cometary nuclei in our previous work \citep{munoz19}. 

In \citetalias{Munoz21}, we divided our set of 34 DPs into three categories according to their characteristic evolution: Regular, Resonant, and Irregular. In this work, we have explored in detail the evolution of the ensemble of 200 realizations of two representative members of the Regular and two of the Resonant families. Specifically, from the regular family, we choose Eris and \dpn{2015}{KH}{162}, and from the resonant family, we choose Pluto and \dpn{2010}{EK}{139}.

The chaotic structure of the TN region results in near identical simulations giving different results in long-term integrations. Thus %using different codes could result in a different evolution for the same object; 
a single simulation is not guaranteed to fall into the most common behavior, or even worse, the real object is not guaranteed to fall into the most common behavior.

The evident advantage of an ensemble of simulations over individual runs is that an ensemble permits exploring an object's potential and most likely evolution, and simultaneously exploring some of its less likely results. Other authors typically consider ensembles of simulations within the observational uncertainty of a particular object \citep[e.g.][]{Bannister16,Volk18,Chen22}, however, in this work, we focus on the dispersion of an ensemble of identical initial conditions whose underlying source is accounted for by time-step differences leading to differences in DP-DP perturbations. In the case of well-known regular and resonant objects (with tiny observational uncertainties), those three sources of dispersion are equivalent in showing the most likely evolution of the object, which in turn can be used to secure the object's classification, mostly when such an object is close to ambiguous locations, like in the case of resonant regions. It is also clear that long-term integrations (over 100 Myr) will show diffusion of the orbital parameters. However, there are permitted limits within these excursions where regular and resonant objects can move, which may not be completely covered in short-term integrations or by single realizations.

%The evident advantage of an ensemble of simulations over individual runs is that an ensemble permits exploring the potential and most likely evolution of an object, and simultaneously exploring some of its less likely results. The most likely evolution can be used to secure the object's classification, mostly when such an object is close to ambiguous locations, like in the case of resonant regions. It is also clear that long-term integrations (over 100 Myr) will show diffusion of the orbital parameters. However, there are permitted limits within these excursions where regular and resonant objects are allowed to move, that may not be completely covered in short-term integrations or by single realizations.

For the regular objects, we found excursions within clear limits in $e$ and $i$ and no divergent diffusion in $a$. Though the frequency evolution of $e$ and the main frequency of $i$ (i.e. the frequency of $\Omega$ and $H$) are different, there is a regular exchange of $e$ and $i$ through the vZLK mechanism. The first of those frequencies is slightly faster, with $e$ oscillating twice faster than $i$ (i.e. the orbital regression of the nodes) in the case of Eris, and three times faster in the case of \dpn{2015}{KH}{162}.

One of the most striking results of this work is the behavior of Eris's $\varpi$. Firstly, we find that at the beginning of our simulations, it has some small oscillations, but it remains locked at approximately $190^{\circ}$; this in itself is not remarkable, interactions with some massive body could force it into a constant angle. However, the required inclination value for this to happen, according to the secular theory of \citet{Gallardo12} (eq. \ref{eqn:varpi}), is too fine-tuned for this to be a coincidence; yet, after $\sim150$ Myr the stability of $\varpi$ is lost. This is not seen in the green curve (third panel of Figure \ref{fig:Eris_angles}); in fact, from this single realization one could argue that there is a nearly constant with a very small regression; however, not all realizations behave the same way and, while all start with a constant $\varpi$, by $\sim150$ Myr the ensemble has lost its coherence. This means that whatever placed Eris into this fixed position is missing from our simulations. This is an indication that some additional perturbation, still present in the current Solar System, has to be considered. Planet 9 or other KB planets \citep[e.g.][]{Batygin19,Lykawka23} come to mind, but any further constraining is out of the scope of this paper. We will explore this idea in future work.

Regarding the evolution of resonant objects, these are much more complex when compared to regular objects, as they are affected by additional perturbations from their resonance. In the case of Pluto, which is locked deeply within the 3:2 MMR, a restoring effect hinders any diffusion in $a$. However, when we study $e$ and $i$, the restoring effect goes much deeper.

If we were to study a single realization or our whole ensemble for less than 150 Myr, we would find a complex orbit with oscillations responding to many frequencies influencing Pluto. However, when plotting the full ensemble for the entire 1 Gyr, we find new additional periods for $e$ and $i$ (as well as for $\omega$ and $H$) where the ensemble rhythmically shifts from being stringently restricted to where it is allowed to have more leeway. 

For all other objects, close encounters with other DPs result in a phase shift (perhaps linked to a small period shift) that will remain until a new encounter occurs; however, for resonant objects, and in particular for Pluto in the 3:2 MMR, these cumulative perturbations will be counterbalanced by the effects of the 3:2 resonance, impeding the possibility of any shift in frequencies and restricting the available phases to only a few values, i.e. sometimes Pluto will return to its original state and other times Pluto's orbit will be forced into the next notch allowed by \pf{P}{e}{p} and \pf{P}{i}{p}. These shifts will occur while retaining the known frequencies of the 3:2 MMR, the vZLK resonance, and the super resonance.

These new periods do not originate directly from Pluto's orbit, but rather from the influence Neptune 

%We also found a $\sim3\%$ difference between the main $e$ and $i$ periods. This difference translates into the Kozai mechanism not holding perfectly, thus being periods when $e$ and $i$ are correlated and periods when they are anti-correlated (as shown in Figure \ref{fig:Pluto_evol}). The interaction and connection between the $e$ and $i$ oscillations, as well as their allowed/restricted rhythms, results in Pluto's orbit having a very rich and complex pattern forced by the frame of allowed regions shown by the full ensemble of realizations. Yet, in all cases $\omega$ remains librating around $90^\circ$ for the whole integration and the whole ensemble (see Figure \ref{fig:Pluto_evol_angles}), showing the extreme stability of Pluto and its persistence within both the MMR and the Kozai resonance.  

We also found a $\sim3\%$ difference between the main $e$ and $i$ periods, \pf{P}{e}{o} and \pf{P}{i}{o}, respectively. This difference translates into periods when $e$ and $i$ are correlated and periods when they are anti-correlated (as shown in Figure \ref{fig:Pluto_evol}). Independently of this, in all cases, $\omega$ remains librating around $90^\circ$ for the whole integration and the whole ensemble (see Figure \ref{fig:Pluto_evol_angles}). The interaction and connection between the $e$ and $i$ oscillations, as well as their allowed/restricted rhythms, results in Pluto's orbit having a very rich and complex pattern forced by the frame of allowed regions shown by the full ensemble of realizations. Yet, the orbit of Pluto is extremely stable, remaining in 3:2 MMR for the whole integration. On the other hand, the vZLK mechanism does not always hold perfectly, but this does not affect Pluto's overall stability. 

Finally, for \dpn{2010}{EK}{139}, while initially trapped in Neptune's 7:2 MMR, it is not as forcefully trapped as Pluto is, partially due to the weaker restoring forces of the 7:2 resonance; this results in \dpn{2010}{EK}{139} likely having a slight diffusion in $a$. Only 20 of our 200 resonances remain well-trapped in the 7:2 MMR, while most of them remain only intermittently within the resonance (and 2 escape altogether). The wide variations of the orbits of \dpn{2010}{EK}{139} are exemplified by the large and irregular changes in $e$ and $i$, which we highlight with two realizations in green and yellow in Figs. \ref{fig:2010EK13_evol} and \ref{fig:2010EK13_angles}. 

We found that a minimum eccentricity of 0.299 is required for \dpn{2010}{EK}{139} to remain librating within the 7:2 MMR. Below this value, circulation ensues, which indicates a narrowing of the resonant region in phase space. 

Though in all realizations there is an interchange of $e$ and $i$, as would be expected from the vZLK mechanism, the required condition for $\omega$ is not fulfilled simultaneously. We found that $\omega$ is most likely to librate around $0^{\circ}$, $180^{\circ}$, or to regress when \dpn{2010}{EK}{139} is well trapped within the 7:2 MMR, while $\omega$ librates around $90^{\circ}$, $270^{\circ}$ or it circulates when the libration of $\phi_{7:2}$ is lost and starts to regress. All of this indicates that \dpn{2010}{EK}{139} is not likely to be trapped simultaneously in MMR and in vZLK resonance, which weakens its overall stability compared to Pluto.

Overall we have shown the necessity of the use of a more realistic model of the outer Solar System, considering the gravitational perturbations from several dozens of smaller objects beyond the GPs. Our more complete system warns about the dangers of using a single orbit and short-term simulations to classify an object, especially those close to MMRs.  Conversely, we have shown the advantage of using a large set of long-term simulations to classify objects based on their most likely evolution. By doing this, we found that Eris's $\varpi$ is tightly confined and remains nearly constant for 150 Myr, but ultimately the simulations lose coherence and end up drifting away. We found that \dpn{2010}{EK}{139} is not well trapped in resonance in the long-term, thus it cannot be confined to MMR and vZLK resonance simultaneously. Finally, for Pluto, we found a new underlying constraining mechanism  associated with Neptune's fundamental secular frequencies $g_8$ and $s_8$, which introduces two new periods that become evident only for an ensemble showing the cumulative effects of close encounters with other DPs.

%We discuss/speculate/argue this is an indication of a missing element in our solar system model, likely an additional planetary perturber. 

\begin{acknowledgments}
The authors acknowledge the anonymous referee for careful reading and insightful comments that helped to improve this paper. MAM acknowledges Universidad de Atacama for the DIUDA grant No. 88231R14. AP acknowledges the DGAPA-PAPIIT grant IG-100622. APV acknowledges the DGAPA-PAPIIT grant IA103122 and IA103224.

\end{acknowledgments}

%\appendix


\bibliography{dpsbib}{}

\begin{thebibliography}{}
\expandafter\ifx\csname natexlab\endcsname\relax\def\natexlab#1{#1}\fi
\providecommand{\url}[1]{\href{#1}{#1}}
\providecommand{\dodoi}[1]{doi:~\href{http://doi.org/#1}{\nolinkurl{#1}}}
\providecommand{\doeprint}[1]{\href{http://ascl.net/#1}{\nolinkurl{http://ascl.net/#1}}}
\providecommand{\doarXiv}[1]{\href{https://arxiv.org/abs/#1}{\nolinkurl{https://arxiv.org/abs/#1}}}

\bibitem[{{Alves-Carmo} {et~al.}(2023){Alves-Carmo}, {Vaillant}, \&
  {Correia}}]{Alves23}
{Alves-Carmo}, A.~J., {Vaillant}, T., \& {Correia}, A. C.~M. 2023, \aap, 677,
  A83, \dodoi{10.1051/0004-6361/202244982}

\bibitem[{{Applegate} {et~al.}(1986){Applegate}, {Douglas}, {Gursel},
  {Sussman}, \& {Wisdom}}]{Applegate86}
{Applegate}, J.~H., {Douglas}, M.~R., {Gursel}, Y., {Sussman}, G.~J., \&
  {Wisdom}, J. 1986, \aj, 92, 176, \dodoi{10.1086/114149}

\bibitem[{{Balaji} {et~al.}(2023){Balaji}, {Zaveri}, {Hayashi}, {Hermosillo
  Ruiz}, {Barnes}, {Murray-Clay}, {Volk}, {Gerhardt}, \& {Syed}}]{Balaji23}
{Balaji}, S., {Zaveri}, N., {Hayashi}, N., {et~al.} 2023, \mnras, 524, 3039,
  \dodoi{10.1093/mnras/stad2026}

\bibitem[{{Bannister} {et~al.}(2016){Bannister}, {Alexandersen}, {Benecchi},
  {Chen}, {Delsanti}, {Fraser}, {Gladman}, {Granvik}, {Grundy},
  {Guilbert-Lepoutre}, {Gwyn}, {Ip}, {Jakubik}, {Jones}, {Kaib}, {Kavelaars},
  {Lacerda}, {Lawler}, {Lehner}, {Lin}, {Lykawka}, {Marsset}, {Murray-Clay},
  {Noll}, {Parker}, {Petit}, {Pike}, {Rousselot}, {Schwamb}, {Shankman},
  {Veres}, {Vernazza}, {Volk}, {Wang}, \& {Weryk}}]{Bannister16}
{Bannister}, M.~T., {Alexandersen}, M., {Benecchi}, S.~D., {et~al.} 2016, \aj,
  152, 212, \dodoi{10.3847/0004-6256/152/6/212}

\bibitem[{{Bannister} {et~al.}(2017){Bannister}, {Shankman}, {Volk}, {Chen},
  {Kaib}, {Gladman}, {Jakubik}, {Kavelaars}, {Fraser}, {Schwamb}, {Petit},
  {Wang}, {Gwyn}, {Alexandersen}, \& {Pike}}]{Bannister17}
{Bannister}, M.~T., {Shankman}, C., {Volk}, K., {et~al.} 2017, \aj, 153, 262,
  \dodoi{10.3847/1538-3881/aa6db5}

\bibitem[{{Batygin} {et~al.}(2019){Batygin}, {Adams}, {Brown}, \&
  {Becker}}]{Batygin19}
{Batygin}, K., {Adams}, F.~C., {Brown}, M.~E., \& {Becker}, J.~C. 2019,
  \physrep, 805, 1, \dodoi{10.1016/j.physrep.2019.01.009}

\bibitem[{{Batygin} {et~al.}(2021){Batygin}, {Mardling}, \&
  {Nesvorn{\'y}}}]{Batygin21}
{Batygin}, K., {Mardling}, R.~A., \& {Nesvorn{\'y}}, D. 2021, \apj, 920, 148,
  \dodoi{10.3847/1538-4357/ac19a4}

\bibitem[{{Carpino} {et~al.}(1987){Carpino}, {Milani}, \& {Nobili}}]{Carpino87}
{Carpino}, M., {Milani}, A., \& {Nobili}, A.~M. 1987, \aap, 181, 182

\bibitem[{{Chambers}(1999)}]{Chambers99}
{Chambers}, J.~E. 1999, \mnras, 304, 793,
  \dodoi{10.1046/j.1365-8711.1999.02379.x}

\bibitem[{{Chen} {et~al.}(2022){Chen}, {Eduardo}, {Mu{\~n}oz-Guti{\'e}rrez},
  {Wang}, {Lehner}, \& {Chang}}]{Chen22}
{Chen}, Y.-T., {Eduardo}, M.~R., {Mu{\~n}oz-Guti{\'e}rrez}, M.~A., {et~al.}
  2022, \apjl, 937, L22, \dodoi{10.3847/2041-8213/ac90b9}

\bibitem[{{Cohen} \& {Hubbard}(1965)}]{Cohen65}
{Cohen}, C.~J., \& {Hubbard}, E.~C. 1965, \aj, 70, 10, \dodoi{10.1086/109674}

\bibitem[{{de Sousa} {et~al.}(2020){de Sousa}, {Morbidelli}, {Raymond},
  {Izidoro}, {Gomes}, \& {Vieira Neto}}]{DeSousa20}
{de Sousa}, R.~R., {Morbidelli}, A., {Raymond}, S.~N., {et~al.} 2020, \icarus,
  339, 113605, \dodoi{10.1016/j.icarus.2019.113605}

\bibitem[{{Dones} {et~al.}(2015){Dones}, {Brasser}, {Kaib}, \&
  {Rickman}}]{Dones15}
{Dones}, L., {Brasser}, R., {Kaib}, N., \& {Rickman}, H. 2015, \ssr, 197, 191,
  \dodoi{10.1007/s11214-015-0223-2}

\bibitem[{{Duncan} {et~al.}(1988){Duncan}, {Quinn}, \& {Tremaine}}]{Duncan88}
{Duncan}, M., {Quinn}, T., \& {Tremaine}, S. 1988, \apjl, 328, L69,
  \dodoi{10.1086/185162}

\bibitem[{{Duncan} {et~al.}(1995){Duncan}, {Levison}, \& {Budd}}]{Duncan95}
{Duncan}, M.~J., {Levison}, H.~F., \& {Budd}, S.~M. 1995, \aj, 110, 3073,
  \dodoi{10.1086/117748}

\bibitem[{{Emel'yanenko} {et~al.}(2004){Emel'yanenko}, {Asher}, \&
  {Bailey}}]{Emelyanenko04}
{Emel'yanenko}, V.~V., {Asher}, D.~J., \& {Bailey}, M.~E. 2004, \mnras, 350,
  161, \dodoi{10.1111/j.1365-2966.2004.07624.x}

\bibitem[{{Fernandez}(1980)}]{Fernandez80}
{Fernandez}, J.~A. 1980, \mnras, 192, 481, \dodoi{10.1093/mnras/192.3.481}

\bibitem[{{Fernandez} \& {Ip}(1984)}]{Fernandez84}
{Fernandez}, J.~A., \& {Ip}, W.~H. 1984, \icarus, 58, 109,
  \dodoi{10.1016/0019-1035(84)90101-5}

\bibitem[{{Forg{\'a}cs-Dajka} {et~al.}(2023){Forg{\'a}cs-Dajka},
  {K{\H{o}}v{\'a}ri}, {Kov{\'a}cs}, {Kiss}, \& {S{\'a}ndor}}]{Forgacs23}
{Forg{\'a}cs-Dajka}, E., {K{\H{o}}v{\'a}ri}, E., {Kov{\'a}cs}, T., {Kiss}, C.,
  \& {S{\'a}ndor}, Z. 2023, \apjs, 266, 5, \dodoi{10.3847/1538-4365/acc4c8}

\bibitem[{{Gallardo} {et~al.}(2012){Gallardo}, {Hugo}, \& {Pais}}]{Gallardo12}
{Gallardo}, T., {Hugo}, G., \& {Pais}, P. 2012, \icarus, 220, 392,
  \dodoi{10.1016/j.icarus.2012.05.025}

\bibitem[{{Gladman} \& {Volk}(2021)}]{Gladman21}
{Gladman}, B., \& {Volk}, K. 2021, \araa, 59, 203,
  \dodoi{10.1146/annurev-astro-120920-010005}

\bibitem[{{Gladman} {et~al.}(2012){Gladman}, {Lawler}, {Petit}, {Kavelaars},
  {Jones}, {Parker}, {Van Laerhoven}, {Nicholson}, {Rousselot}, {Bieryla}, \&
  {Ashby}}]{Gladman12}
{Gladman}, B., {Lawler}, S.~M., {Petit}, J.-M., {et~al.} 2012, \aj, 144, 23,
  \dodoi{10.1088/0004-6256/144/1/23}

\bibitem[{{Hahn}(2003)}]{Hahn03}
{Hahn}, J.~M. 2003, \apj, 595, 531, \dodoi{10.1086/377195}

\bibitem[{{Huang} {et~al.}(2022){Huang}, {Gladman}, \& {Volk}}]{Huang22}
{Huang}, Y., {Gladman}, B., \& {Volk}, K. 2022, \apjs, 259, 54,
  \dodoi{10.3847/1538-4365/ac559a}

\bibitem[{{Ida} {et~al.}(2000){Ida}, {Bryden}, {Lin}, \& {Tanaka}}]{Ida00}
{Ida}, S., {Bryden}, G., {Lin}, D.~N.~C., \& {Tanaka}, H. 2000, \apj, 534, 428,
  \dodoi{10.1086/308720}

\bibitem[{{Ito} \& {Ohtsuka}(2019)}]{Ito19}
{Ito}, T., \& {Ohtsuka}, K. 2019, Monographs on Environment, Earth and Planets,
  7, 1, \dodoi{10.5047/meep.2019.00701.0001}

\bibitem[{{Ito} \& {Tanikawa}(2002)}]{Ito02}
{Ito}, T., \& {Tanikawa}, K. 2002, \mnras, 336, 483,
  \dodoi{10.1046/j.1365-8711.2002.05765.x}

\bibitem[{{Kaib} \& {Sheppard}(2016)}]{Kaib16}
{Kaib}, N.~A., \& {Sheppard}, S.~S. 2016, \aj, 152, 133,
  \dodoi{10.3847/0004-6256/152/5/133}

\bibitem[{{Khain} {et~al.}(2020){Khain}, {Becker}, {Lin}, {Gerdes}, {Adams},
  {Bernardinelli}, {Bernstein}, {Franson}, {Markwardt}, {Hamilton}, {Napier},
  {Sako}, {Abbott}, {Avila}, {Bertin}, {Brooks}, {Buckley-Geer}, {Burke},
  {Carnero Rosell}, {Carrasco Kind}, {Carretero}, {Costa}, {Vicente}, {Desai},
  {Diehl}, {Doel}, {Flaugher}, {Frieman}, {Garc{\'\i}a-Bellido}, {Gaztanaga},
  {Gruen}, {Gruendl}, {Gschwend}, {Gutierrez}, {Hollowood}, {Honscheid},
  {James}, {Kuropatkin}, {Maia}, {Marshall}, {Menanteau}, {Miller}, {Miquel},
  {Plazas}, {Sanchez}, {Scarpine}, {Schubnell}, {Sevilla-Noarbe}, {Smith},
  {Sobreira}, {Suchyta}, {Swanson}, {Tarle}, {Walker}, {Wester}, \& {Dark
  Energy Survey Collaboration}}]{Khain20}
{Khain}, T., {Becker}, J.~C., {Lin}, H.~W., {et~al.} 2020, \aj, 159, 133,
  \dodoi{10.3847/1538-3881/ab7002}

\bibitem[{{Kinoshita} \& {Nakai}(1996)}]{Kinoshita96}
{Kinoshita}, H., \& {Nakai}, H. 1996, Earth Moon and Planets, 72, 165,
  \dodoi{10.1007/BF00117514}

\bibitem[{{Knezevic} {et~al.}(1991){Knezevic}, {Milani}, {Farinella},
  {Froeschle}, \& {Froeschle}}]{Knezevic91}
{Knezevic}, Z., {Milani}, A., {Farinella}, P., {Froeschle}, C., \& {Froeschle},
  C. 1991, \icarus, 93, 316, \dodoi{10.1016/0019-1035(91)90215-F}

\bibitem[{{Laskar}(1988)}]{Laskar88}
{Laskar}, J. 1988, \aap, 198, 341

\bibitem[{{Levison} \& {Duncan}(1997)}]{Levison97}
{Levison}, H.~F., \& {Duncan}, M.~J. 1997, \icarus, 127, 13,
  \dodoi{10.1006/icar.1996.5637}

\bibitem[{{Li} {et~al.}(2019){Li}, {Xia}, \& {Zhou}}]{Li2019}
{Li}, J., {Xia}, Z.~J., \& {Zhou}, L. 2019, \aap, 630, A68,
  \dodoi{10.1051/0004-6361/201834196}

\bibitem[{{Lykawka} \& {Ito}(2023)}]{Lykawka23}
{Lykawka}, P.~S., \& {Ito}, T. 2023, \aj, 166, 118,
  \dodoi{10.3847/1538-3881/aceaf0}

\bibitem[{{Lykawka} \& {Mukai}(2005)}]{Lykawka05}
{Lykawka}, P.~S., \& {Mukai}, T. 2005, Earth Moon and Planets, 97, 107,
  \dodoi{10.1007/s11038-005-9056-4}

\bibitem[{{Malhotra}(1995)}]{Malhotra95}
{Malhotra}, R. 1995, \aj, 110, 420, \dodoi{10.1086/117532}

\bibitem[{{Malhotra} {et~al.}(2000){Malhotra}, {Duncan}, \&
  {Levison}}]{Malhotra00}
{Malhotra}, R., {Duncan}, M.~J., \& {Levison}, H.~F. 2000, in Protostars and
  Planets IV, ed. V.~{Mannings}, A.~P. {Boss}, \& S.~S. {Russell}, 1231,
  \dodoi{10.48550/arXiv.astro-ph/9901155}

\bibitem[{{Malhotra} \& {Ito}(2022)}]{Malhotra22}
{Malhotra}, R., \& {Ito}, T. 2022, Proceedings of the National Academy of
  Science, 119, 2118692119, \dodoi{10.1073/pnas.2118692119}

\bibitem[{{Malhotra} {et~al.}(2018){Malhotra}, {Lan}, {Volk}, \&
  {Wang}}]{Malhotra18}
{Malhotra}, R., {Lan}, L., {Volk}, K., \& {Wang}, X. 2018, \aj, 156, 55,
  \dodoi{10.3847/1538-3881/aac9c3}

\bibitem[{{Malhotra} \& {Williams}(1997)}]{Malhotra97}
{Malhotra}, R., \& {Williams}, J.~G. 1997, in Pluto and Charon, ed. S.~A.
  {Stern} \& D.~J. {Tholen}, 127

\bibitem[{{Matheson} \& {Malhotra}(2023)}]{Matheson23}
{Matheson}, I., \& {Malhotra}, R. 2023, \aj, 165, 241,
  \dodoi{10.3847/1538-3881/accffd}

\bibitem[{{Milani} \& {Nobili}(1992)}]{Milani92}
{Milani}, A., \& {Nobili}, A.~M. 1992, \nat, 357, 569, \dodoi{10.1038/357569a0}

\bibitem[{{Milani} {et~al.}(1989){Milani}, {Nobili}, \& {Carpino}}]{Milani89}
{Milani}, A., {Nobili}, A.~M., \& {Carpino}, M. 1989, \icarus, 82, 200,
  \dodoi{10.1016/0019-1035(89)90031-6}

\bibitem[{{Morbidelli} {et~al.}(2008){Morbidelli}, {Levison}, \&
  {Gomes}}]{Morbidelli08}
{Morbidelli}, A., {Levison}, H.~F., \& {Gomes}, R. 2008, in The Solar System
  Beyond Neptune, ed. M.~A. {Barucci}, H.~{Boehnhardt}, D.~P. {Cruikshank},
  A.~{Morbidelli}, \& R.~{Dotson}, 275--292,
  \dodoi{10.48550/arXiv.astro-ph/0703558}

\bibitem[{{Morbidelli} {et~al.}(1995){Morbidelli}, {Thomas}, \&
  {Moons}}]{Morbidelli95}
{Morbidelli}, A., {Thomas}, F., \& {Moons}, M. 1995, \icarus, 118, 322,
  \dodoi{10.1006/icar.1995.1194}

\bibitem[{{Mu{\~n}oz-Guti{\'e}rrez} {et~al.}(2021){Mu{\~n}oz-Guti{\'e}rrez},
  {Peimbert}, {Lehner}, \& {Wang}}]{Munoz21}
{Mu{\~n}oz-Guti{\'e}rrez}, M.~A., {Peimbert}, A., {Lehner}, M.~J., \& {Wang},
  S.-Y. 2021, \aj, 162, 164, \dodoi{10.3847/1538-3881/ac1102}

\bibitem[{{Mu{\~n}oz-Guti{\'e}rrez} {et~al.}(2019){Mu{\~n}oz-Guti{\'e}rrez},
  {Peimbert}, {Pichardo}, {Lehner}, \& {Wang}}]{Munoz19}
{Mu{\~n}oz-Guti{\'e}rrez}, M.~A., {Peimbert}, A., {Pichardo}, B., {Lehner},
  M.~J., \& {Wang}, S.~Y. 2019, \aj, 158, 184, \dodoi{10.3847/1538-3881/ab4399}

\bibitem[{{Murray} \& {Dermott}(1999)}]{Murray99}
{Murray}, C.~D., \& {Dermott}, S.~F. 1999, {Solar System Dynamics},
  \dodoi{10.1017/CBO9781139174817}

\bibitem[{{Nacozy} \& {Diehl}(1978)}]{Nacozy78}
{Nacozy}, P.~E., \& {Diehl}, R.~E. 1978, \aj, 83, 522, \dodoi{10.1086/112231}

\bibitem[{{Nesvorn{\'y}}(2018)}]{Nesvorny18}
{Nesvorn{\'y}}, D. 2018, \araa, 56, 137,
  \dodoi{10.1146/annurev-astro-081817-052028}

\bibitem[{{Nesvorn{\'y}} \& {Morbidelli}(2012)}]{Nesvorny12}
{Nesvorn{\'y}}, D., \& {Morbidelli}, A. 2012, \aj, 144, 117,
  \dodoi{10.1088/0004-6256/144/4/117}

\bibitem[{{Nesvorn{\'y}} {et~al.}(2017){Nesvorn{\'y}}, {Vokrouhlick{\'y}},
  {Dones}, {Levison}, {Kaib}, \& {Morbidelli}}]{Nesvorny17}
{Nesvorn{\'y}}, D., {Vokrouhlick{\'y}}, D., {Dones}, L., {et~al.} 2017, \apj,
  845, 27, \dodoi{10.3847/1538-4357/aa7cf6}

\bibitem[{{P{\'a}l} {et~al.}(2012){P{\'a}l}, {Kiss}, {M{\"u}ller},
  {Santos-Sanz}, {Vilenius}, {Szalai}, {Mommert}, {Lellouch}, {Rengel},
  {Hartogh}, {Protopapa}, {Stansberry}, {Ortiz}, {Duffard}, {Thirouin},
  {Henry}, \& {Delsanti}}]{Pal12}
{P{\'a}l}, A., {Kiss}, C., {M{\"u}ller}, T.~G., {et~al.} 2012, \aap, 541, L6,
  \dodoi{10.1051/0004-6361/201218874}

\bibitem[{{Petit} {et~al.}(2011){Petit}, {Kavelaars}, {Gladman}, {Jones},
  {Parker}, {Van Laerhoven}, {Nicholson}, {Mars}, {Rousselot}, {Mousis},
  {Marsden}, {Bieryla}, {Taylor}, {Ashby}, {Benavidez}, {Campo Bagatin}, \&
  {Bernabeu}}]{Petit11}
{Petit}, J.-M., {Kavelaars}, J.~J., {Gladman}, B.~J., {et~al.} 2011, \aj, 142,
  131, \dodoi{10.1088/0004-6256/142/4/131}

\bibitem[{{Porter} {et~al.}(2018){Porter}, {Buie}, {Parker}, {Spencer},
  {Benecchi}, {Tanga}, {Verbiscer}, {Kavelaars}, {Gwyn}, {Young}, {Weaver},
  {Olkin}, {Parker}, \& {Stern}}]{Porter18}
{Porter}, S.~B., {Buie}, M.~W., {Parker}, A.~H., {et~al.} 2018, \aj, 156, 20,
  \dodoi{10.3847/1538-3881/aac2e1}

\bibitem[{{Robutel} \& {Laskar}(2001)}]{Robutel01}
{Robutel}, P., \& {Laskar}, J. 2001, \icarus, 152, 4,
  \dodoi{10.1006/icar.2000.6576}

\bibitem[{{Saillenfest}(2020)}]{Saillenfest20}
{Saillenfest}, M. 2020, Celestial Mechanics and Dynamical Astronomy, 132, 12,
  \dodoi{10.1007/s10569-020-9954-9}

\bibitem[{{Saillenfest} \& {Lari}(2017)}]{Saillenfest17}
{Saillenfest}, M., \& {Lari}, G. 2017, \aap, 603, A79,
  \dodoi{10.1051/0004-6361/201730525}

\bibitem[{{Sheppard} {et~al.}(2016){Sheppard}, {Trujillo}, \&
  {Tholen}}]{Sheppard16}
{Sheppard}, S.~S., {Trujillo}, C., \& {Tholen}, D.~J. 2016, \apjl, 825, L13,
  \dodoi{10.3847/2041-8205/825/1/L13}

\bibitem[{{Souami} \& {Souchay}(2012)}]{Souami12}
{Souami}, D., \& {Souchay}, J. 2012, \aap, 543, A133,
  \dodoi{10.1051/0004-6361/201219011}

\bibitem[{{Sussman} \& {Wisdom}(1988)}]{Sussman88}
{Sussman}, G.~J., \& {Wisdom}, J. 1988, Science, 241, 433,
  \dodoi{10.1126/science.241.4864.433}

\bibitem[{{Thommes} {et~al.}(1999){Thommes}, {Duncan}, \&
  {Levison}}]{Thommes99}
{Thommes}, E.~W., {Duncan}, M.~J., \& {Levison}, H.~F. 1999, \nat, 402, 635,
  \dodoi{10.1038/45185}

\bibitem[{{Torbett}(1989)}]{Torbett89}
{Torbett}, M.~V. 1989, \aj, 98, 1477, \dodoi{10.1086/115233}

\bibitem[{{Tsiganis} {et~al.}(2005){Tsiganis}, {Gomes}, {Morbidelli}, \&
  {Levison}}]{Tsiganis05}
{Tsiganis}, K., {Gomes}, R., {Morbidelli}, A., \& {Levison}, H.~F. 2005, \nat,
  435, 459, \dodoi{10.1038/nature03539}

\bibitem[{{Volk} {et~al.}(2018){Volk}, {Murray-Clay}, {Gladman}, {Lawler},
  {Yu}, {Alexandersen}, {Bannister}, {Chen}, {Dawson}, {Greenstreet}, {Gwyn},
  {Kavelaars}, {Lin}, {Lykawka}, \& {Petit}}]{Volk18}
{Volk}, K., {Murray-Clay}, R.~A., {Gladman}, B.~J., {et~al.} 2018, \aj, 155,
  260, \dodoi{10.3847/1538-3881/aac268}

\bibitem[{{Wan} {et~al.}(2001){Wan}, {Huang}, \& {Innanen}}]{Wan01}
{Wan}, X.~S., {Huang}, T.~Y., \& {Innanen}, K.~A. 2001, \aj, 121, 1155,
  \dodoi{10.1086/318733}

\bibitem[{{Williams} \& {Benson}(1971)}]{Williams71}
{Williams}, J.~G., \& {Benson}, G.~S. 1971, \aj, 76, 167,
  \dodoi{10.1086/111100}

\end{thebibliography}
\bibliographystyle{aasjournal}



\end{document}